\documentclass[12pt]{article}
\usepackage{epsfig}
\usepackage{amsmath}
\usepackage{hhline}
\usepackage{amssymb}
\usepackage{times}
\usepackage{cite}

\newlength{\dinwidth}
\newlength{\dinmargin}
\begin{document}  

\begin{titlepage}

\begin{flushright}
IEKP-KA/2001-24 \\
Nov. 2001
\end{flushright}

\vspace*{1.7cm}

\begin{center}
\begin{LARGE}
{\bf Proton and Photon Structure \footnote{Invited talk at the 
XX International Symposium on Lepton and Photon Interactions at High Energies,
Rome, Italy, July 2001}}
\end{LARGE}

\vspace{0.8cm}

{\large Martin Erdmann} \\

\vspace{0.8cm}

\noindent
Institut f\"ur Experimentelle Kernphysik,
Universit\"at Karlsruhe, \\ Wolfgang-Gaede-Str. 1, 
D-76131 Karlsruhe \\
Martin.Erdmann@cern.ch

\end{center}

\vspace{1.2cm}

\begin{abstract}
The increasing precision of the measurements on the proton structure
and an improved treatment of the correlated systematic
experimental errors constitute a major step forward in our understanding
of the flavour decomposition of the proton and the momentum distributions
of the various flavours.
Together with theoretical progress on the next-to-next-to-leading
order QCD corrections to deep inelastic scattering processes,
the proton measurements already imply a new level of precision
for the strong coupling constant $\alpha_s$.
The progress in the measurements on the quantum
fluctuations of the photon allows questions on 
the universal properties of hadronic structures to be addressed.
\end{abstract}

\end{titlepage}

\section{Introduction}

\begin{figure}[htb]
\setlength{\unitlength}{1.0cm}
\begin{picture}(6.0,3.5)
\put(1,0){\epsfig{file=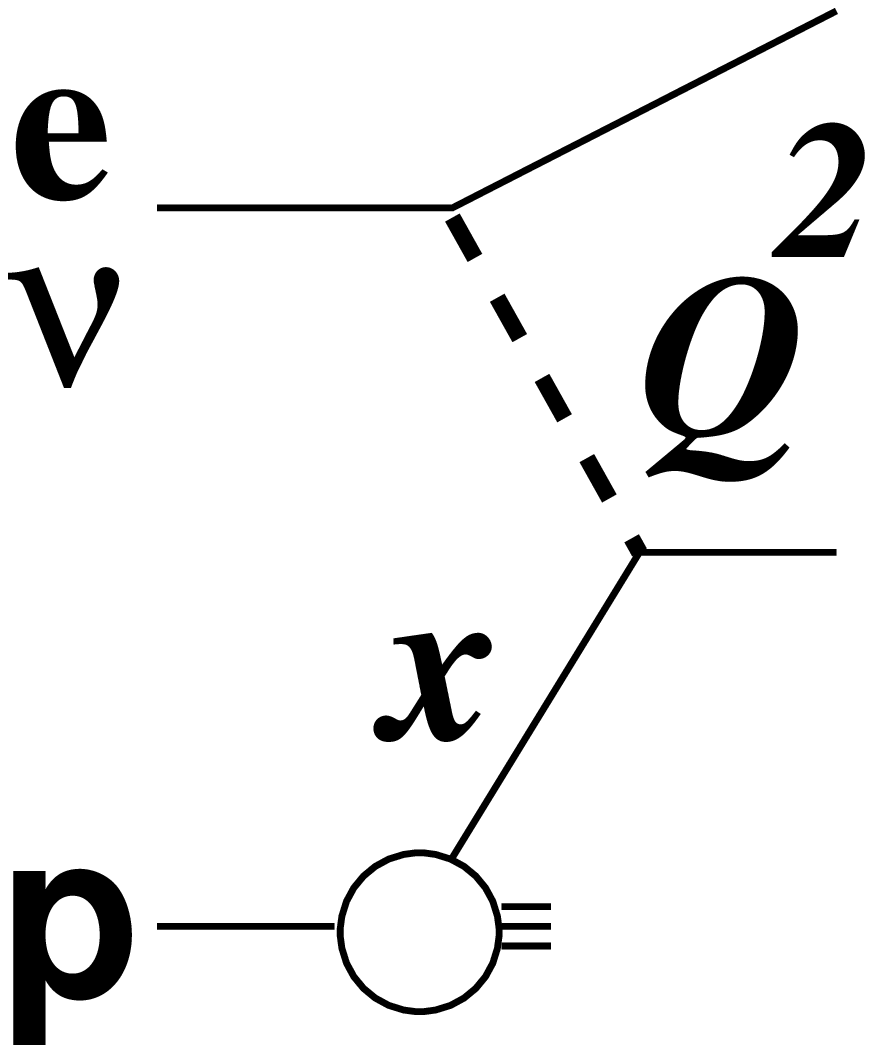,width=2.5cm}}
\put(6.5,0){\epsfig{file=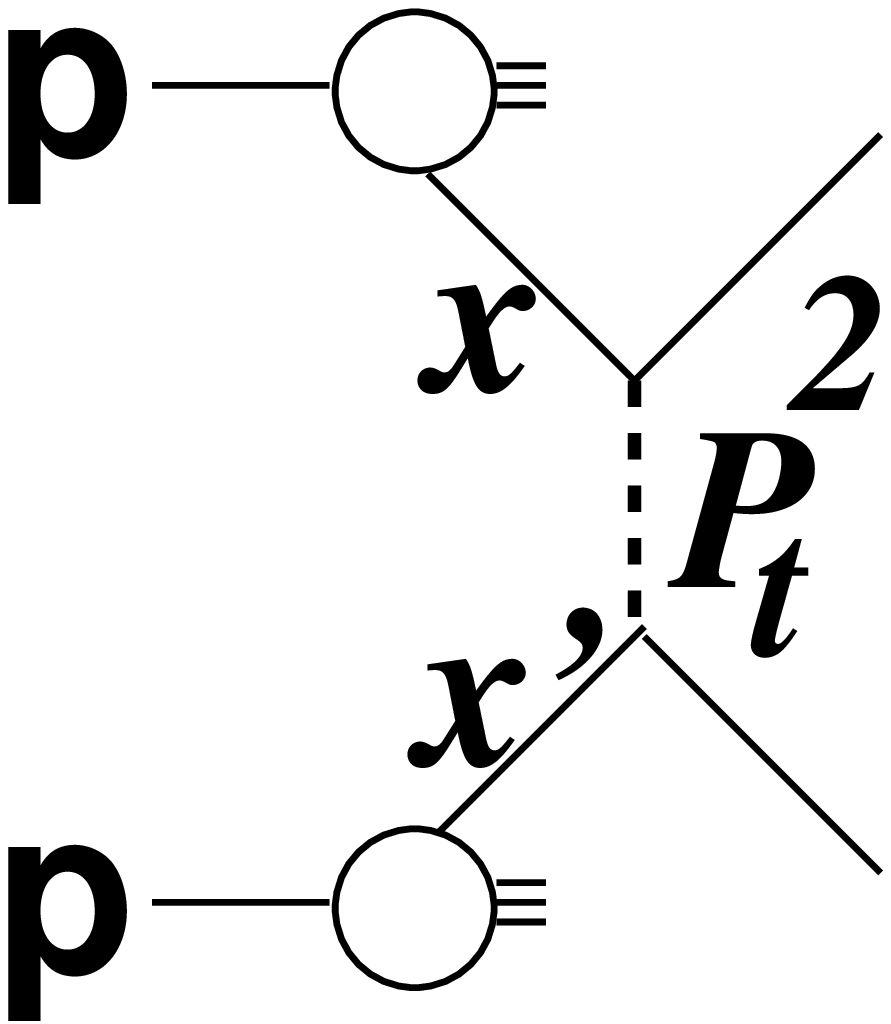,width=2.6cm}}
\put(12,0.1){\epsfig{file=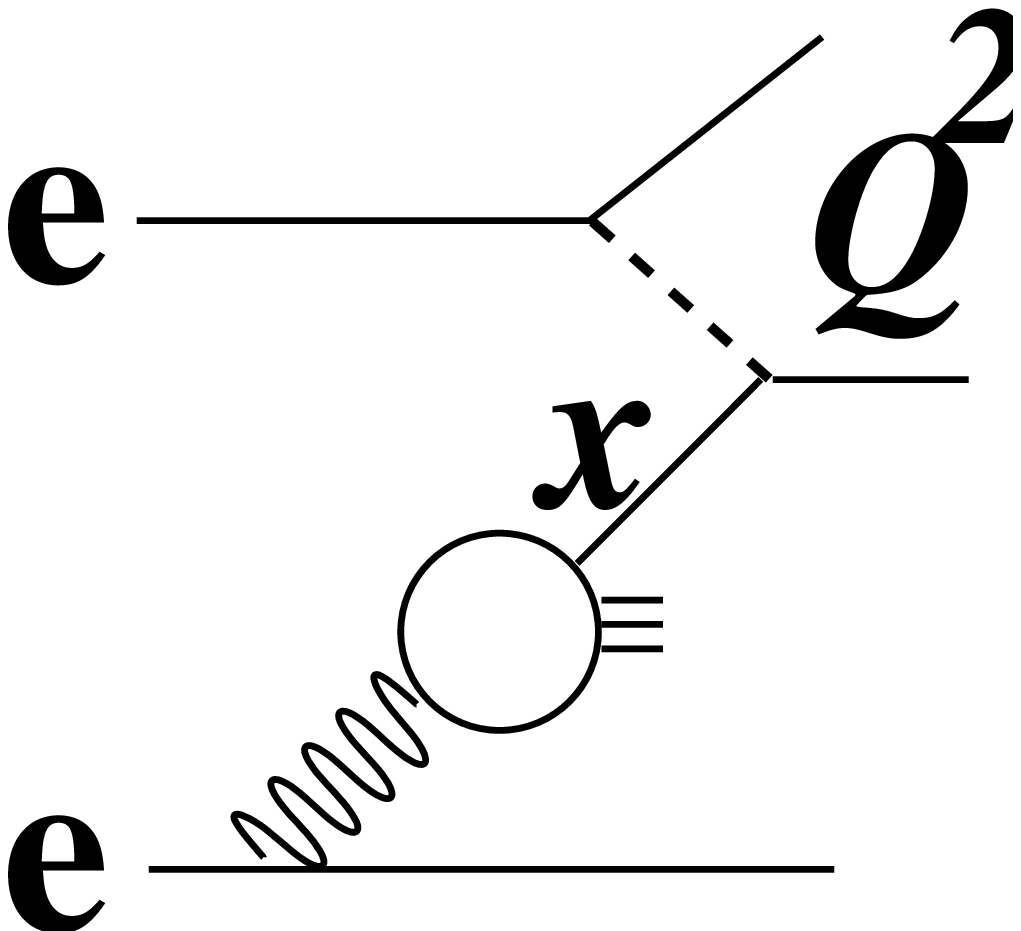,width=3cm}}
\end{picture}
\caption{\label{fig:process} Examples of lepton--proton, proton--proton,
and electron--photon scattering processes probing the proton and photon
structure }
\label{fig:processes}
\end{figure}

\noindent
The research goals for the measurements of the proton and the photon structure
are mainly twofold.
One motivation is the physics understanding of hadronic structures 
concerning the momentum distributions of their partons and the quark flavours 
that contribute.
This aspect is thoroughly analysed in different scattering processes with the 
proton, and in measurements of the hadronic structures developing from the
quantum fluctuations of the photon
(Fig.~\ref{fig:processes}).

The second aim is the test of the Standard Model in both the electroweak and
the strong interaction sector.
The new precision measurements of the proton structure function together
with a new level of higher order QCD corrections results in most precise
determinations of the strong coupling constant.
Furthermore, unification of the electromagnetic and weak interactions is 
under direct observation by the recent deep inelastic scattering data.
The distributions of observables related to the electroweak lepton--quark 
interactions are thoroughly scanned for possible new interactions and 
new heavy resonances.

\section{The Proton}

\begin{figure}[htb]
\setlength{\unitlength}{1cm}
\begin{picture}(16.0,10)
\put(3,0){\epsfig{file=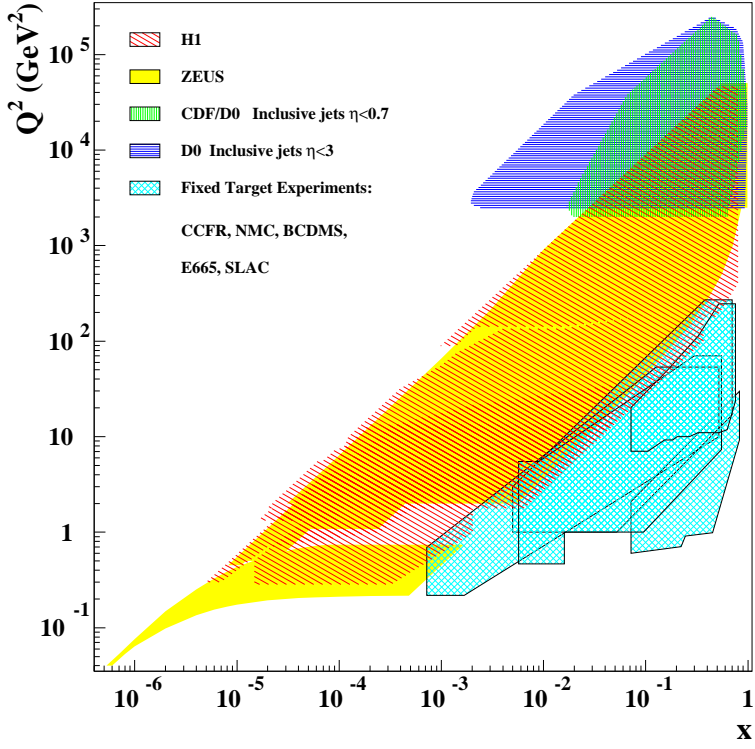,width=10cm}}
\end{picture}
\caption{ Kinematic plane }
\label{fig:kinematic}
\end{figure}

\noindent
Two observables are required for the investigation of the proton structure:
the Bjorken fractional momentum $x$ of the quark relative to the proton momentum
(Fig.~\ref{fig:processes}),
and the resolution scale.
In the case of deep inelastic lepton--proton
scattering, the resolution scale is related to the negative squared four-momentum transfer $Q^2$, 
and for hadron--hadron scattering processes, it is 
related to the squared transverse 
momentum $P_t^2$ produced in the hard collision.
In Fig.~\ref{fig:kinematic}, 
the currently covered kinematic regions are shown,
reaching in $x$ from $10^{-6}$ to almost $1$, and in $Q^2$ ($P_t^2$) from below 
$1$ up to $10^5\,$GeV$^2$.
The latter values allow structures between $1$ fm down
to $1$~am ($10^{-18}$m) to be resolved.

The new results are presented along a triangular walk through the kinematic plane.
We start with the new HERA measurements at high $Q^2$, give the new results of the 
HERA high statistics region at lower $Q^2$, report on the new FNAL fixed target 
results in the same $Q^2$ region at higher values of $x$, and return to the
high $Q^2$ regime with the new TEVATRON di-jet data.

\subsection{Deep inelastic electron--proton scattering}

\subsubsection{Neutral and charged current cross sections}

\noindent
The double differential cross sections for neutral and
charged current lepton--proton scattering are given by
eqs. ~(\ref{eq:nc}) and ~(\ref{eq:cc}).
\begin{eqnarray}
\frac{{\rm d}^2\sigma_{NC}}{{\rm d}Q^2\, {\rm d}x} \; &=& \;2\pi\;\;\;
\alpha^2 \;\;\;\;\;\;\;\;\;\;\;\; \frac{1}{Q^4} \;\;\;\;\;\;\;\;\;\,
 \frac{1}{x} \; \Phi_{NC}( x, Q^2 ) \label{eq:nc}\\
\frac{{\rm d}^2\sigma_{CC}}{{\rm d}Q^2\, {\rm d}x} \; &=& \;\frac{1}{2\pi}\;\;
G_F^2 \; \left(\frac{M_W^2}{M_W^2+Q^2}\right)^2 \; \frac{1}{x} 
\; \Phi_{CC}( x, Q^2 )  \label{eq:cc}
\end{eqnarray}
Here $\alpha$ and $G_F$ are the electromagnetic and Fermi coupling constants,
respectively,
$M_W$ denotes the $W$ boson mass, and the $\Phi$ terms
contain the spin dependencies, the quark flavours, and,
in the case of neutral current scattering, also the 
contributions of $Z$ boson exchange.

Equations ~(\ref{eq:lophincpm}--\ref{eq:lophiccm})
show approximate expressions for the $\Phi$ terms resulting from 
the helicity dependence ($y=Q^2/x/s_{ep}$) of the lepton--quark interactions 
and the prevailing flavours at high ~$x$.
\begin{eqnarray}
\Phi_{NC}^{l\pm N}( x, Q^2 )  &\sim& \left( 1+(1-y)^2 \right) 
\left[\frac{4}{9} xu(x,Q^2) + \frac{1}{9} xd(x,Q^2) + ...\right] 
\label{eq:lophincpm}
\\
\Phi_{CC}^{l+ N}( x, Q^2 ) &\sim& x\bar{u}(x,Q^2) + (1-y)^2 \; xd(x,Q^2)  
\label{eq:lophiccp}
\\
\Phi_{CC}^{l- N}( x, Q^2 ) &\sim& xu(x,Q^2) + (1-y)^2 \; x\bar{d}(x,Q^2) 
\label{eq:lophiccm}
\end{eqnarray}

\subsubsection{Unification of the electroweak interactions} 

\begin{figure}[htb]
\setlength{\unitlength}{1cm}
\begin{picture}(16.0,10)
\put(4.5,-0.7){\epsfig{file=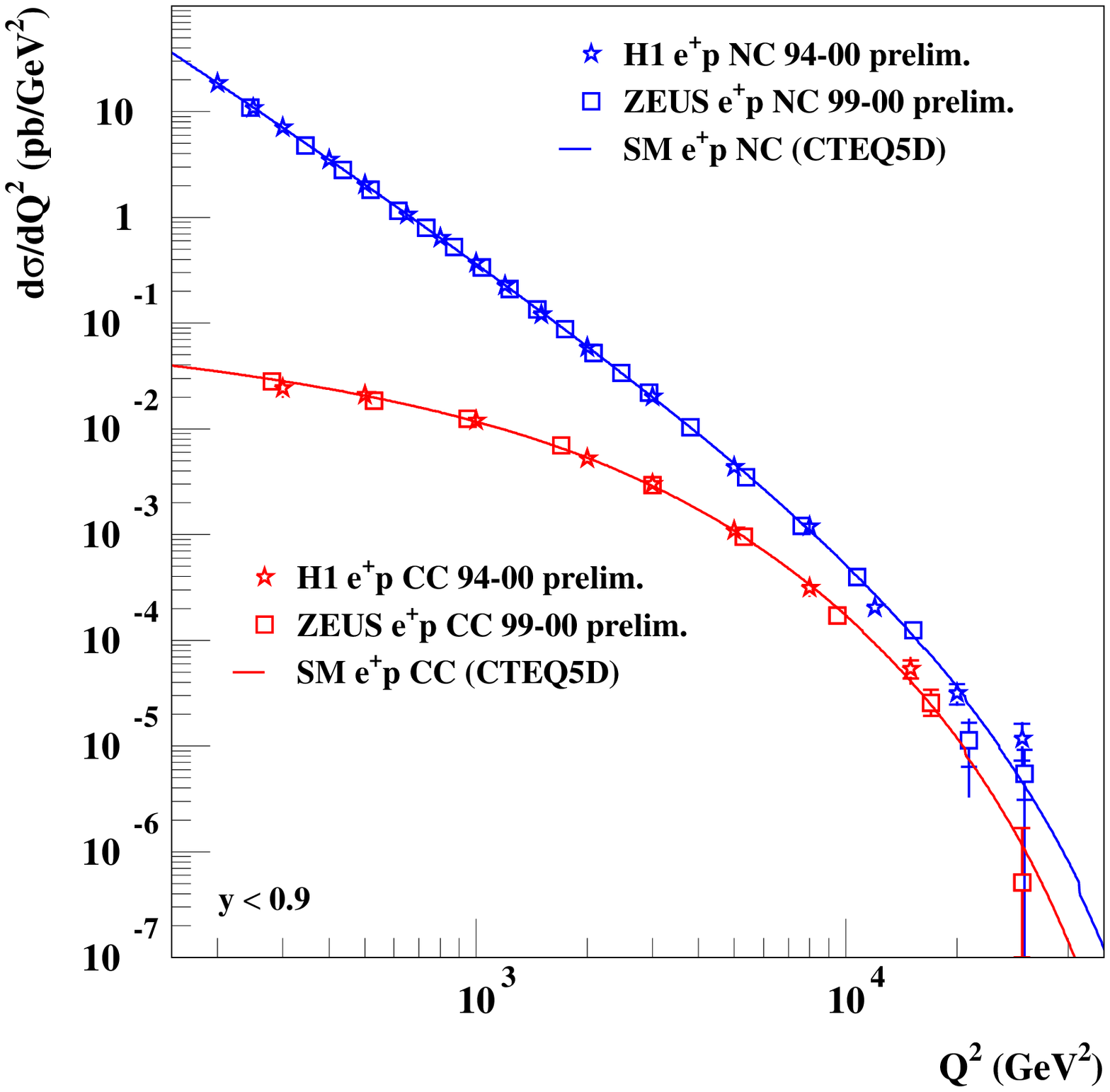,width=11cm}}
\put(0.5,5){\epsfig{file=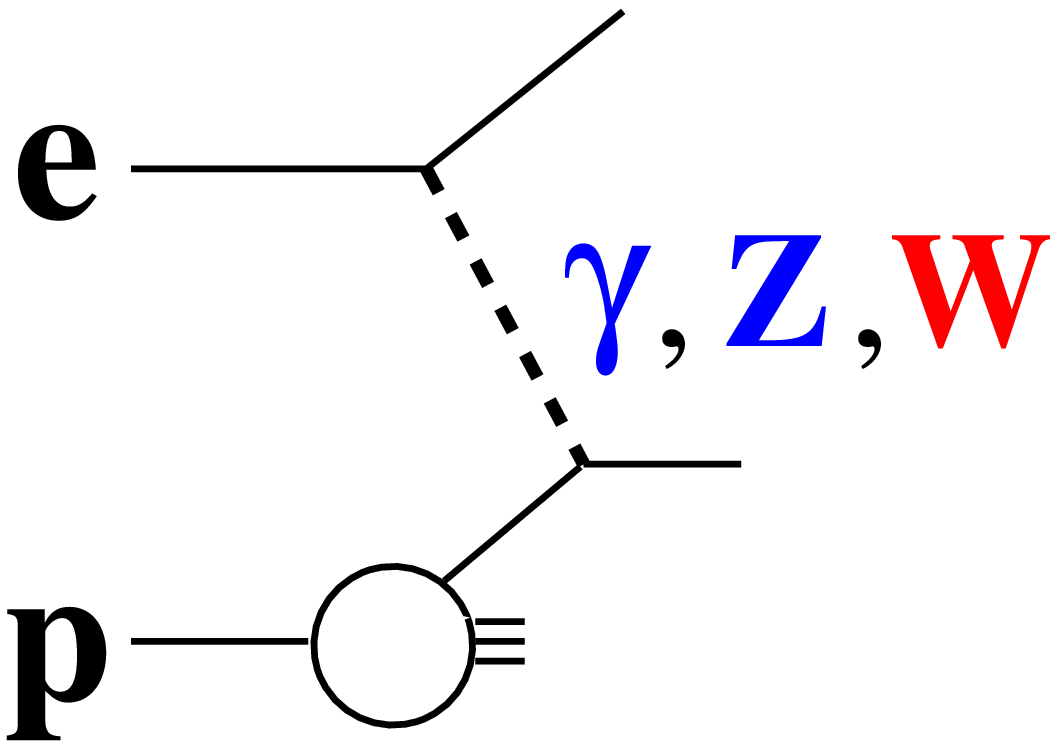,width=3cm}} 
\end{picture}
\caption{Neutral and charged current cross
sections in positron--proton scattering }
\label{fig:unification} 
\end{figure}

\noindent
The large center of mass energy of the HERA collider 
($\sqrt{s_{ep}}=320\,$GeV) allows the neutral and charged 
current interactions to be measured within a single experiment.
In Fig.~\ref{fig:unification}, measurements\cite{hiq2h1,hiq2zeus}
of the neutral and charged 
current cross sections in positron--proton scattering are directly 
compared in terms of single differential cross sections 
as a function of $Q^2$.

The neutral current data fall approximately according to
$Q^{-4}$ as expected from photon exchange.
The charged current measurements exhibit a weak $Q^2$ dependence
at $Q^2$ values of a few hundred GeV$^2$ due to the large $W$ mass.
When $Q^2$ is of the order of the heavy boson masses squared, both
cross sections are measured to be of similar magnitude.

This is a rather direct demonstration of the unification of
the electromagnetic and weak interactions.
Note that this is an approximate statement, since different 
quark flavours of the proton are probed in $e^+p$ scattering, 
the data are integrated over a large
range of $x$, and the photon--$Z$ interference modifies the
cross section for large $Q^2$.

\subsubsection{Helicity dependence and valence quarks}

\noindent
Both HERA experiments have excellent measurements of the 
event kinematics which enables searches for new phenomena
by scanning, e.g., the angular distributions of the lepton--quark
($eq$) interactions
in bins of the $eq$ center of mass energy.

\begin{figure}
\setlength{\unitlength}{1cm}
\begin{picture}(16.0,6.5)
\put(6.8,5.5){\large a)}
\put(14.7,5.5){\large b)}
\put(2,3.9){\epsfig{file=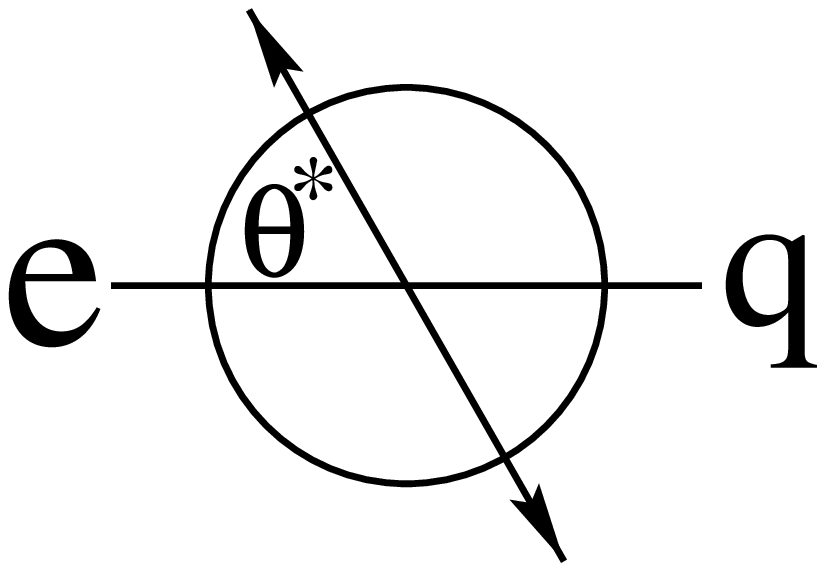,width=2.5cm}}
\put(15.6,3.2){\Large ${\rm u}$}
\put(7.7,2.7){\Large ${\rm d}$}
\put(7.7,1.65){\Large $\bar{\rm u}$}
\put(8.0,0){\epsfig{file=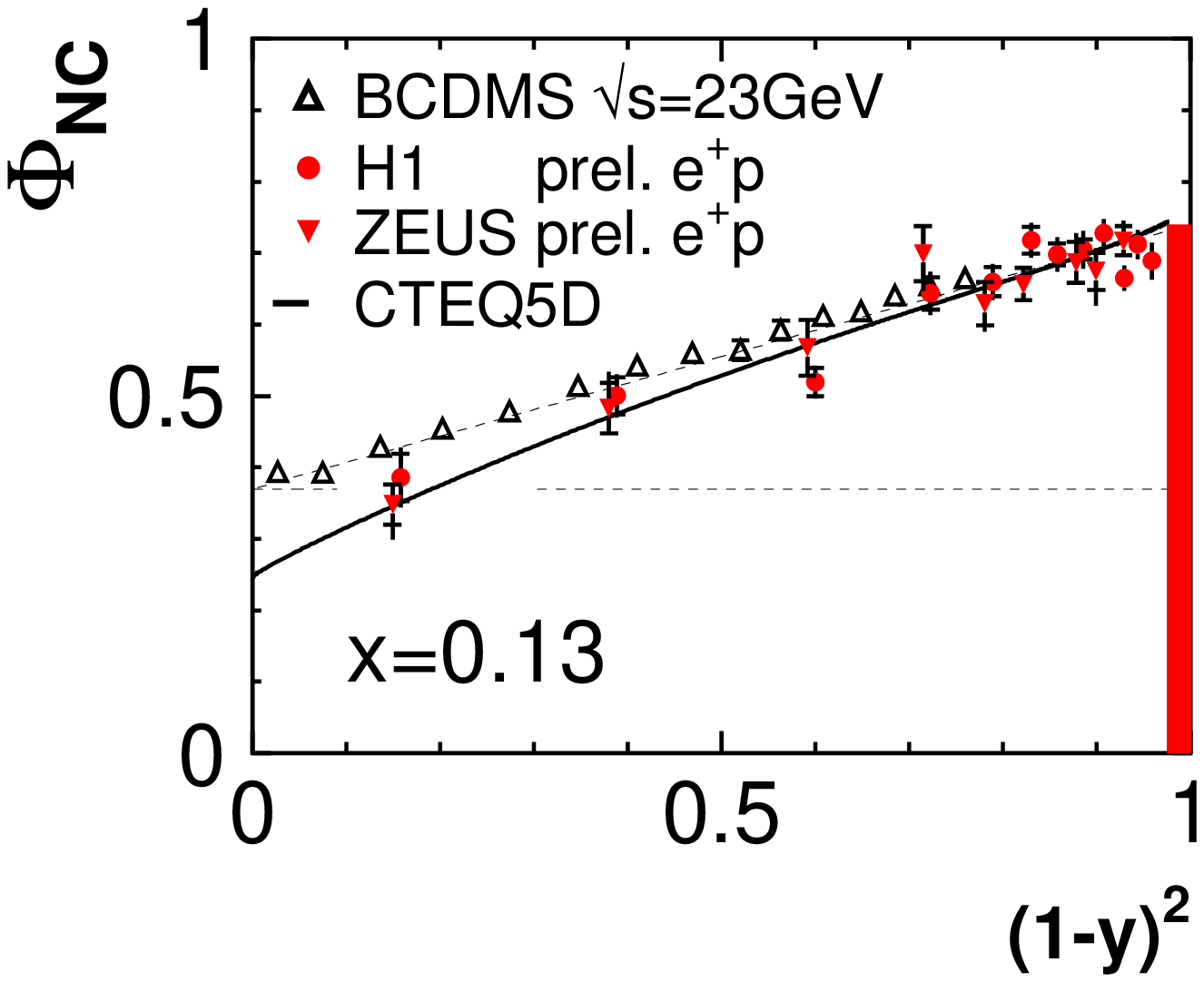,width=8.3cm}}
\put(0,0){\epsfig{file=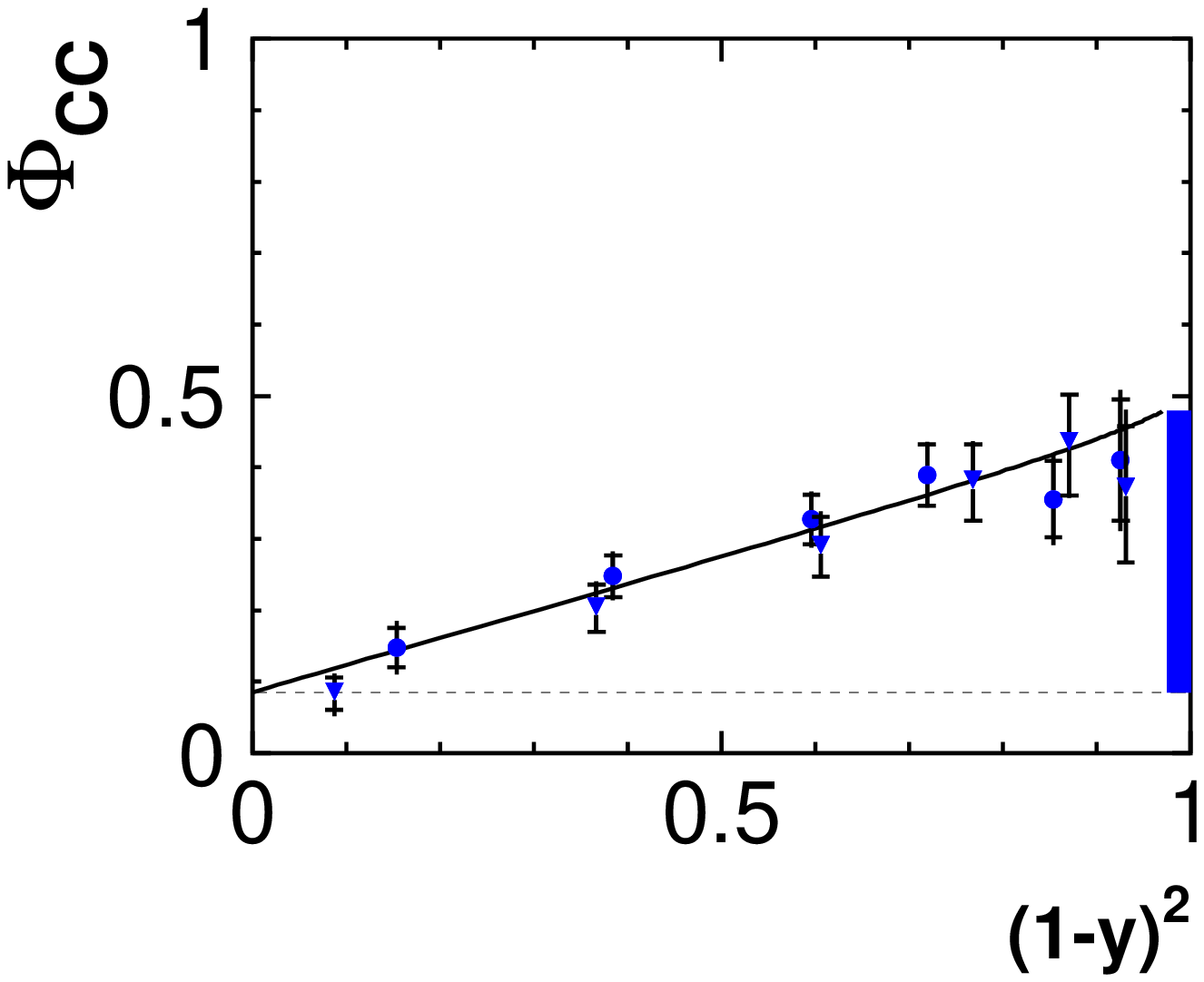,width=8.3cm}}
\end{picture}
\caption{ Helicity weighted parton distributions in charged and neutral
current interactions }
\label{fig:uud}
\end{figure}

The helicity states of the lepton and the quark lead to two
components in the distribution of the center of mass scattering angle 
$\theta^*$ (insert of Fig.~\ref{fig:uud}a).
If both incoming particles have the same handedness, any scattering
angle is allowed and the cross section angular dependent
weight of this component is unity.
If the lepton and the quark have opposite handedness, the total spin of
$1$ cannot be flipped.
Backward scattering is forbidden, 
and the cross section contribution is weighted by
$\cos^4{(\theta^*/2)}=(1-y)^2$.

When analysing charged current interactions in positron--proton collisions,
the positron couples only to negatively charged quarks, e.g. $d, \bar{u}$
(see eq.~(\ref{eq:lophiccp})).
In addition, the right-handed positrons 
couple only to right-handed anti-quarks, and to left-handed quarks.
In Fig.~\ref{fig:uud}a, the $\Phi_{CC}$ term of eq.~(\ref{eq:cc}) is shown
as a function of $(1-y)^2$.
$\Phi_{CC}$ has been obtained by dividing the measured double differential
cross section \cite{hiq2h1,hiq2zeus}
by the constant term of eq.~(\ref{eq:cc}), the Fermi coupling, 
the propagator term, and by multiplying by $x$ to get quark momentum 
distributions.
Here $x=0.13$ has been chosen so as 
to minimise effects of gluon radiation processes.
Within the errors of the measurements, the data are well compatible with
a linearly rising distribution, showing the constant anti-quark 
contribution in the backward scattering region by the intercept
(($1-y)^2\rightarrow 0$).
The quark contribution causes the rising component and is, at this value 
of $x$, mainly that of the $d$ valence quark.

In Fig.~\ref{fig:uud}b, the corresponding neutral current measurement is
shown.
The open triangles show $\Phi_{NC}$ for the BCDMS data \cite{bcdms}
at small center of mass energies ($\sqrt{s_{\mu p}}$ up to $23\,$GeV).
These data follow a linear rise where the intercept and the slope
are of equal magnitude.
This is expected for photon exchange (see eq.~(\ref{eq:lophincpm})).
Owing to the charge dependence of the electromagnetic interactions, these
data are mainly sensitive to the $u$ valence quark distribution at the 
given value of $x$.
Both components of the angular distribution are weighted with $e_u^2=4/9$
such that in the forward scattering region ($(1-y)^2\rightarrow 1$)
an approximate measure of the total $u$ valence quark density is obtained.

The HERA data, which probe the proton at much higher resolution scales,
show the same $u$ quark density in the forward scattering region.
The measurements in the backward scattering region
($(1-y)^2\rightarrow 0$) show the tendency to
be below the BCDMS data which results from negative interference between
the photon and the $Z$ boson exchange in positron--proton scattering.
The smaller cross section is confirmed in the neighbouring $x$-bins
(not shown in the figure).

The direct comparison of the neutral current and charged current measurements
reveals that the proton -- also when probed with the high resolution scales at HERA --
shows the familiar $uud$ valence structure.

No significant sign of new physics has been observed in the inclusive 
measurements so far.
An excess of the data over the Standard Model in the backward scattering 
region at $x\sim 0.4$, previously reported for the $e^+p$ data 
\cite{highq2h1,highq2zeus}, 
has not been confirmed with much larger integrated luminosity \cite{h1lq}
and is hence attributed to a statistical fluctuation.

The HERA upgrade programme has already started and 
is expected to provide a factor of $10$ more integrated luminosity
than obtained so far.
It will provide much higher precision for the
flavour decomposition of the proton as well as for the consistency checks
with respect to the Standard Model predictions.

\subsubsection{\boldmath Measurements of the structure function $F_2$}

\noindent
In the region $1<Q^2\sim 100\,$GeV$^2$, each of the HERA experiments 
has several
million events which are used to provide about $100$ cross section 
measurements.
In the high-statistics region, the accuracy is $2-3$\% which is dominated by
systematic uncertainties \cite{f2h1,f2zeus}.
The measurements are expressed in terms of the 
electromagnetic proton structure function $F_2$ for photon exchange alone
which is related to $\Phi_{NC}$ of eq.~(\ref{eq:nc}) in the way described below.

$\Phi_{NC}$ is commonly written in terms of the generalized structure functions
$\tilde{F_2}, \tilde{F_L}, \tilde{F_3}$ shown in eq. ~(\ref{eq:phinc}).
$\tilde{F_2}$ is 
related to the sum of the quark- and anti-quark distributions, and
$\tilde{F_3}$ to their difference.
$\tilde{F_L}$ describes the proton structure when probed by a longitudinal
photon.
$Y_\pm=1\pm (1-y)^2$ give the helicity dependence of the lepton--quark scattering.

$F_2$ is contained in $\tilde{F_2}$ through eq.~(\ref{eq:f2}),
where $F_2^Z$  and $F_2^{\gamma Z}$  are the contributions due to 
$Z$ exchange and $\gamma Z$ interference, respectively,
$M_Z$ is the mass of the $Z$-boson, \\
\mbox{$\kappa_w=1/(4 \sin^2{\theta_w} \cos^2{\theta_w})$} 
where $\theta_w$ is the Weinberg angle,
and $v$ and $a$ are the vector and axial vector 
couplings of the electron to the $Z$. 
At small $Q^2$ and small $y$, 
the familiar relation (\ref{eq:f2approx}) holds to a good approximation.
\begin{eqnarray}
\Phi_{NC}^{l\pm N}( x, Q^2 ) &=& Y_+ \; \tilde{F_2}(x,Q^2) \; - y^2 \; 
\tilde{F_L}(x,Q^2) \; 
\mp \; Y_- \; x \tilde{F_3}(x,Q^2)
\label{eq:phinc}\\
\tilde{F_2} &=& F_2 - v \frac{\kappa_w Q^2}{(Q^2 + M_Z^2)} F_2^{\gamma Z}
+ (v^2+a^2) \left(\frac{\kappa_w Q^2}{Q^2 + M_Z^2}\right)^2 F_2^Z
\label{eq:f2}\\
\Phi_{NC} ( x, Q^2 ) &\sim& \left( 1 + (1-y)^2 \right) \; F_2 ( x, Q^2 ) \;\;\;\;\;
[{\rm for}\; Q^2<1000 \;{\rm GeV}^2, y<0.3]
\label{eq:f2approx}
\end{eqnarray}

\begin{figure}
\setlength{\unitlength}{1cm}
\begin{picture}(6.0,10.5)
\put(0,0){\epsfig{file=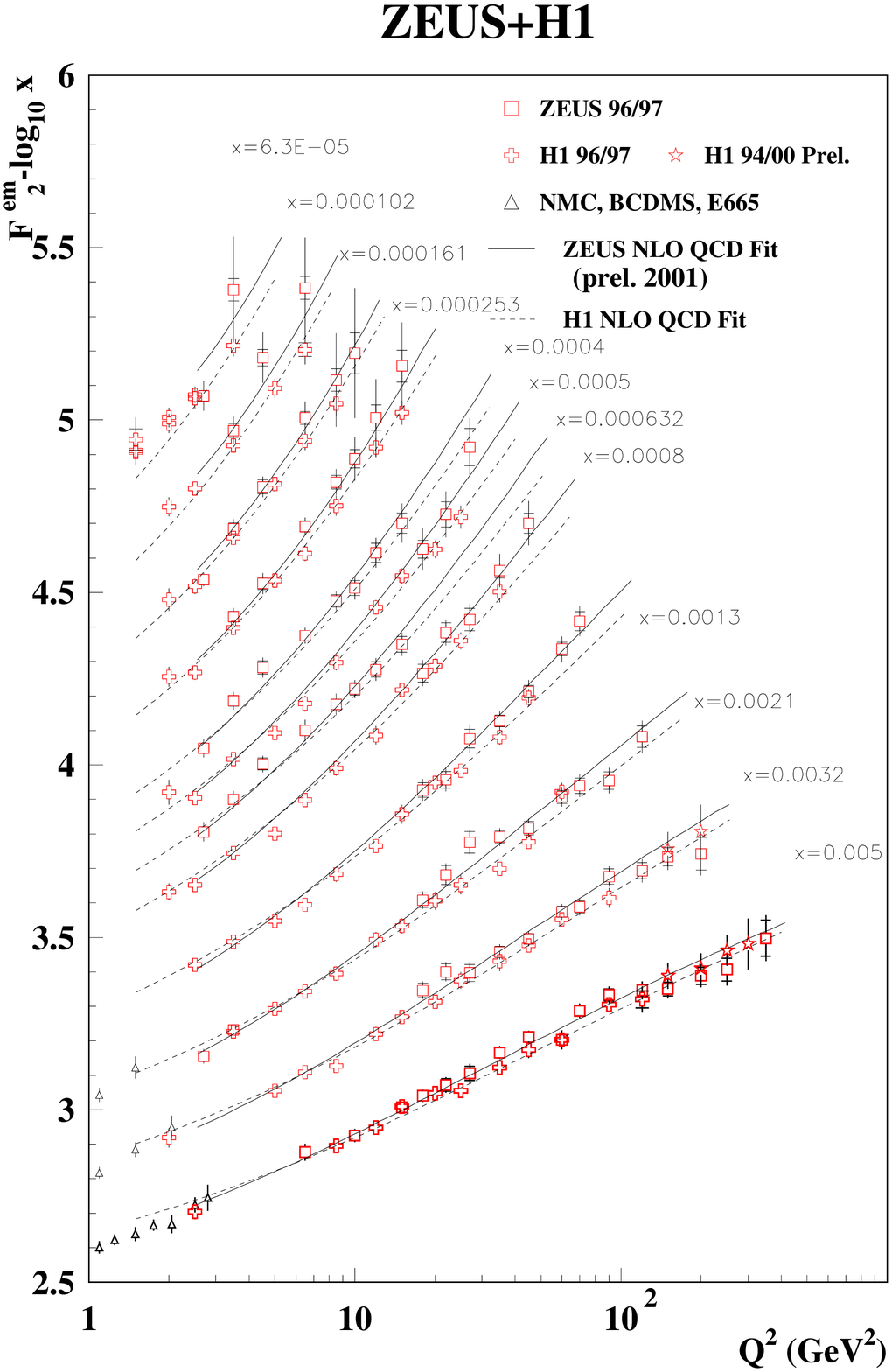,width=7.6cm}}
\put(8.5,0){\epsfig{file=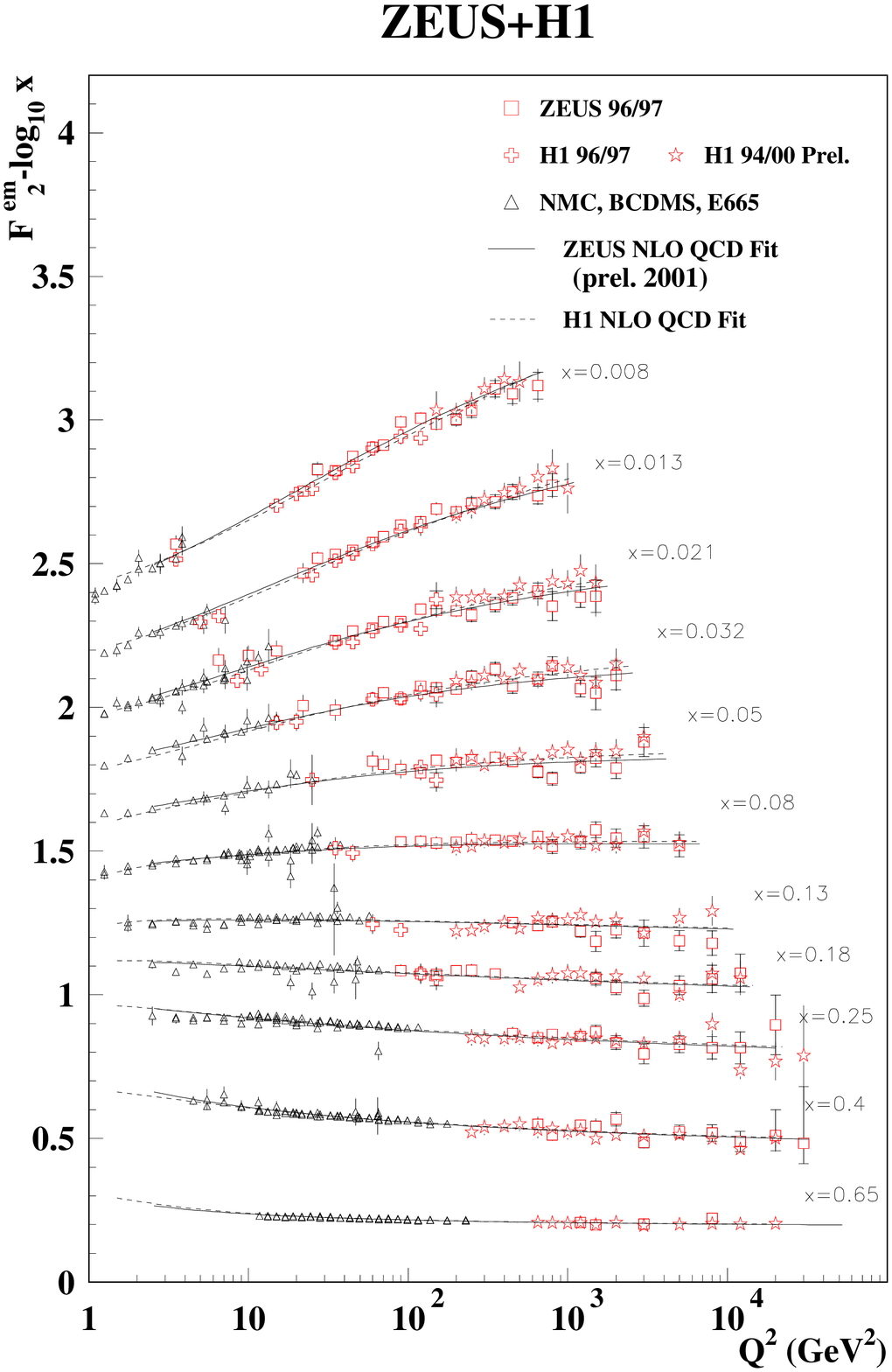,width=7.6cm}}
\end{picture}
\caption{ Proton structure function $F_2$ }
\label{fig:proton-f2}
\end{figure}

In Fig.~\ref{fig:proton-f2}, the $F_2$ measurements are shown in bins of $x$
as a function of $Q^2$.
The measurements are corrected for electroweak radiative effects.
The data show strong scaling violations as expected from QCD radiative effects.

To provide an intuitive understanding of the hadronic 
structure as given by the data, we summarize the data at fixed 
$x$ values and for $2\le Q^2\le 200$ GeV$^2$ with two parameters 
$a(x)$ and $\kappa(x)$ \cite{f2martin}:
\begin{equation}
F_2(x,Q^2) = a(x) \; \left[ \ln{\left(\frac{Q^2}{\Lambda^2}\right)} 
\right]^{\;\; \kappa(x)} \; .
\label{eq:f2me}
\end{equation}
The scale $\Lambda$ is chosen to be a typical value of the
strong interaction scale $\Lambda=350\,$MeV.
The parameter $\kappa$ reflects the strength of the scaling violations,
and $a$ represents the quark distribution at $Q^2=0.3\,$GeV$^2$.
Here the logarithmic term of (\ref{eq:f2me}) is $1$ and therefore $F_2$
is independent of $\kappa$.
The data are well described by the fits using eq.~(\ref{eq:f2me})
(not shown here).

\begin{figure}
\setlength{\unitlength}{1cm}
\begin{picture}(16.0,9.0)
\put(2.5,8.2){\LARGE sea}
\put(5.7,8.2){\LARGE valence}
\put(-0.5,0){\epsfig{file=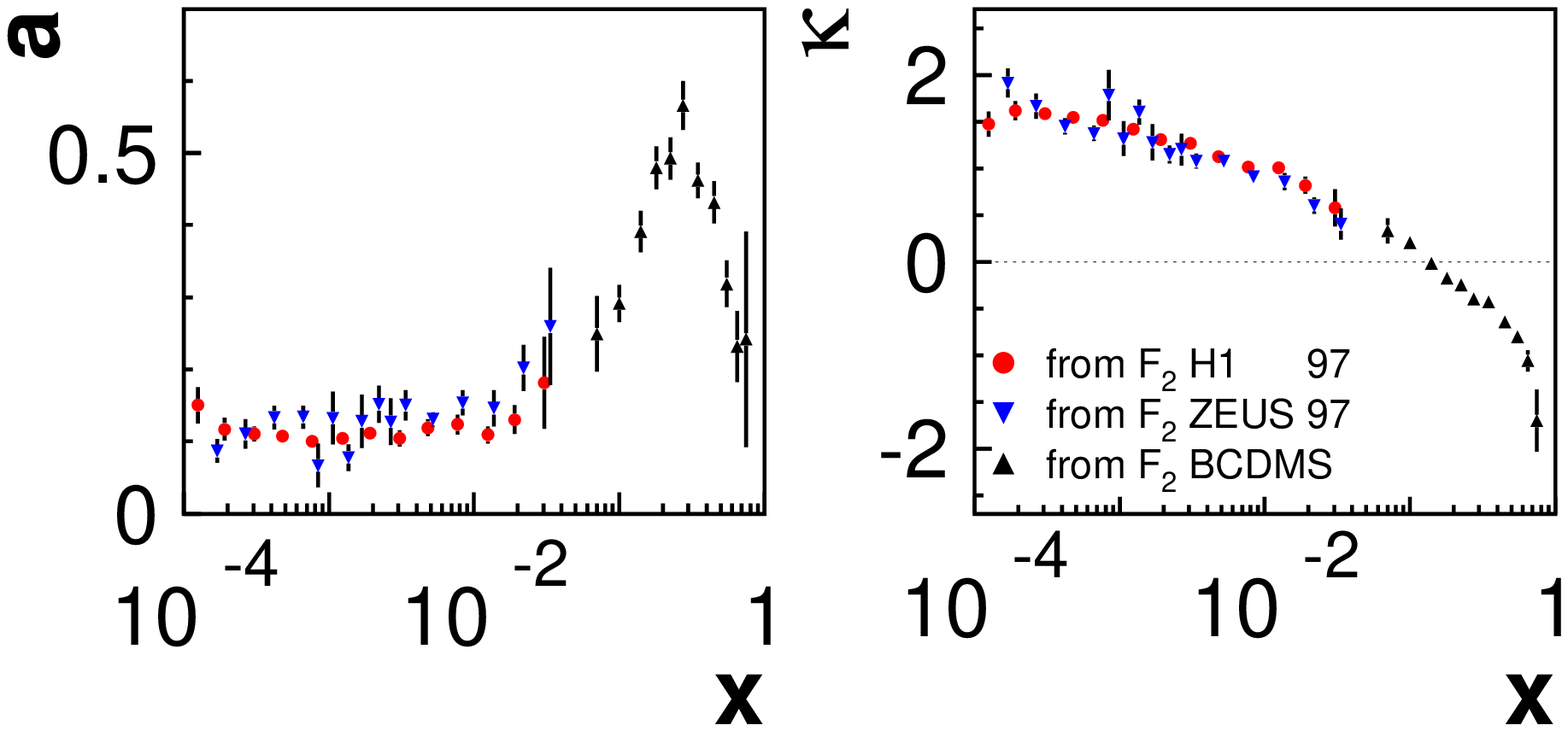,width=17cm}}
\put(14.3,8){\epsfig{file=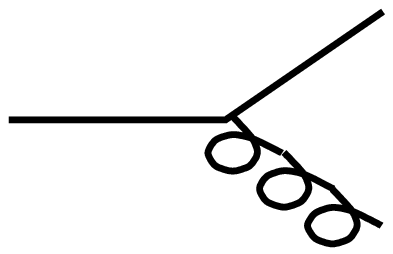,width=1.cm}}
\put(10.5,8){\epsfig{file=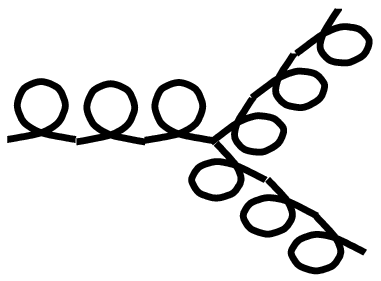,width=0.8cm}}
\end{picture}
\caption{Quark distribution $a(x)$ of the proton at $Q^2=0.3$ GeV$^2$
and scaling violations $\kappa(x)$ }
\label{fig:proton-ak} 
\end{figure}

The fit results for the parameter $a$ are presented in Fig.~\ref{fig:proton-ak}.
The BCDMS data show the valence quarks around the value $x\sim 1/3$.
The HERA data access the low-$x$ region and reveal a sea quark distribution which
is consistent with being constant in $x$.

Around $x=0.1$, the parameter $\kappa$ is zero such that $F_2$ scales.
Large negative scaling violations are visible above $x=0.1$, 
and positive scaling violations
below that value which appear to rise towards small $x$.
With the current data precision it remains an open question whether
or not the scaling violations keep rising towards small $x$ which 
may give information on upper bounds for parton densities.

Some differences in the strength of the scaling violations $\kappa(x)$
between the measurements of the H1 and ZEUS experiments 
are apparent in $a(x)$ which need to be clarified.

The QCD radiative effects causing the scaling violations give 
access to the gluon distribution and the strong coupling constant $\alpha_s$.
Since the splitting functions dominating the radiative effects depend on the 
region of $x$ (see examples of the splitting functions at the top of
Fig.~\ref{fig:proton-ak}), 
the gluon distribution and $\alpha_s$ can be determined simultaneously.

\begin{figure}[htb]
\setlength{\unitlength}{1cm}
\begin{picture}(6.0,8.5)
\put(0,7.8){\large a)}
\put(8.3,7.8){\large b)}
\put(0.5,0){\epsfig{file=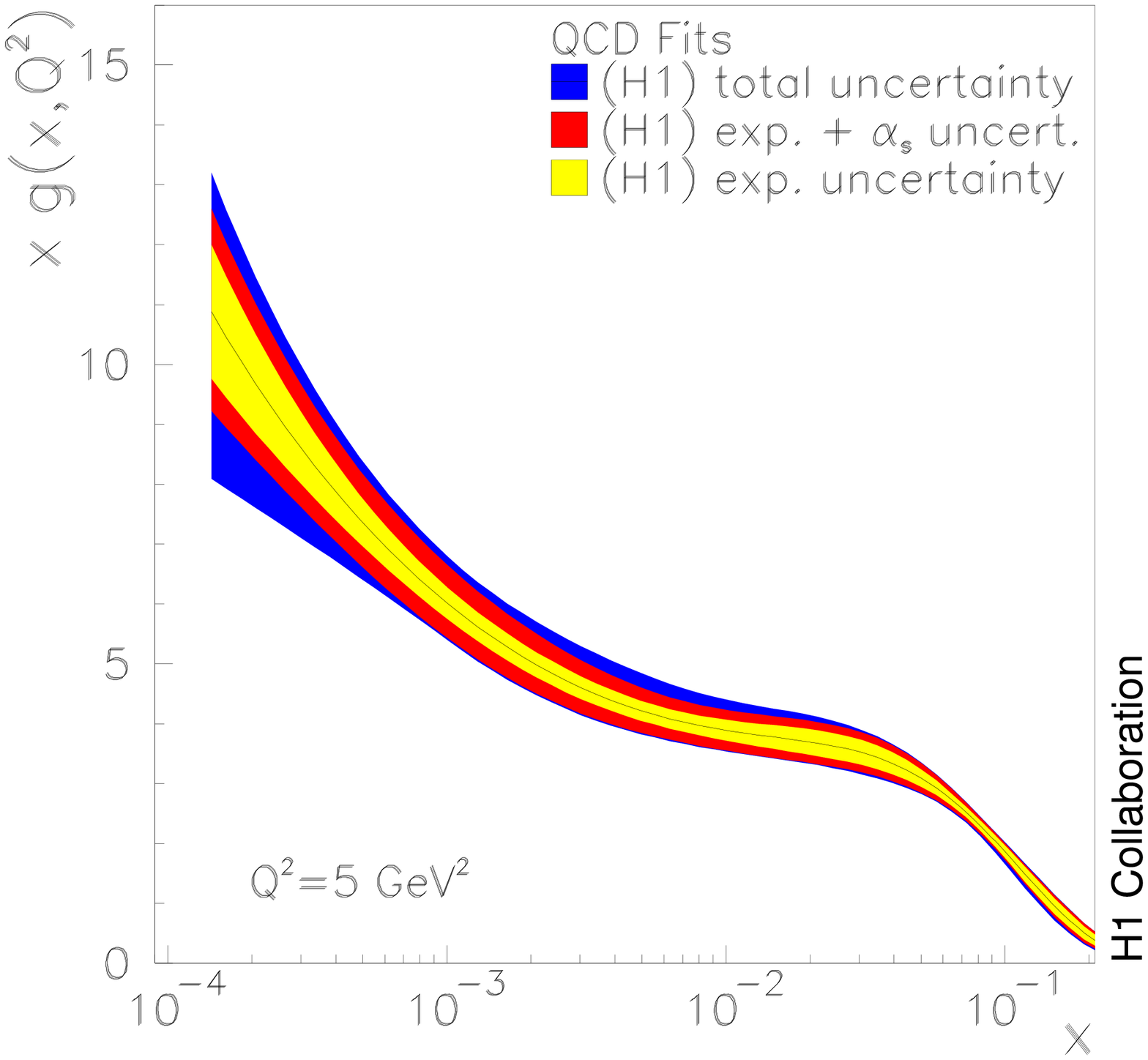,width=7.8cm}}
\put(8.8,-0.2){\epsfig{file=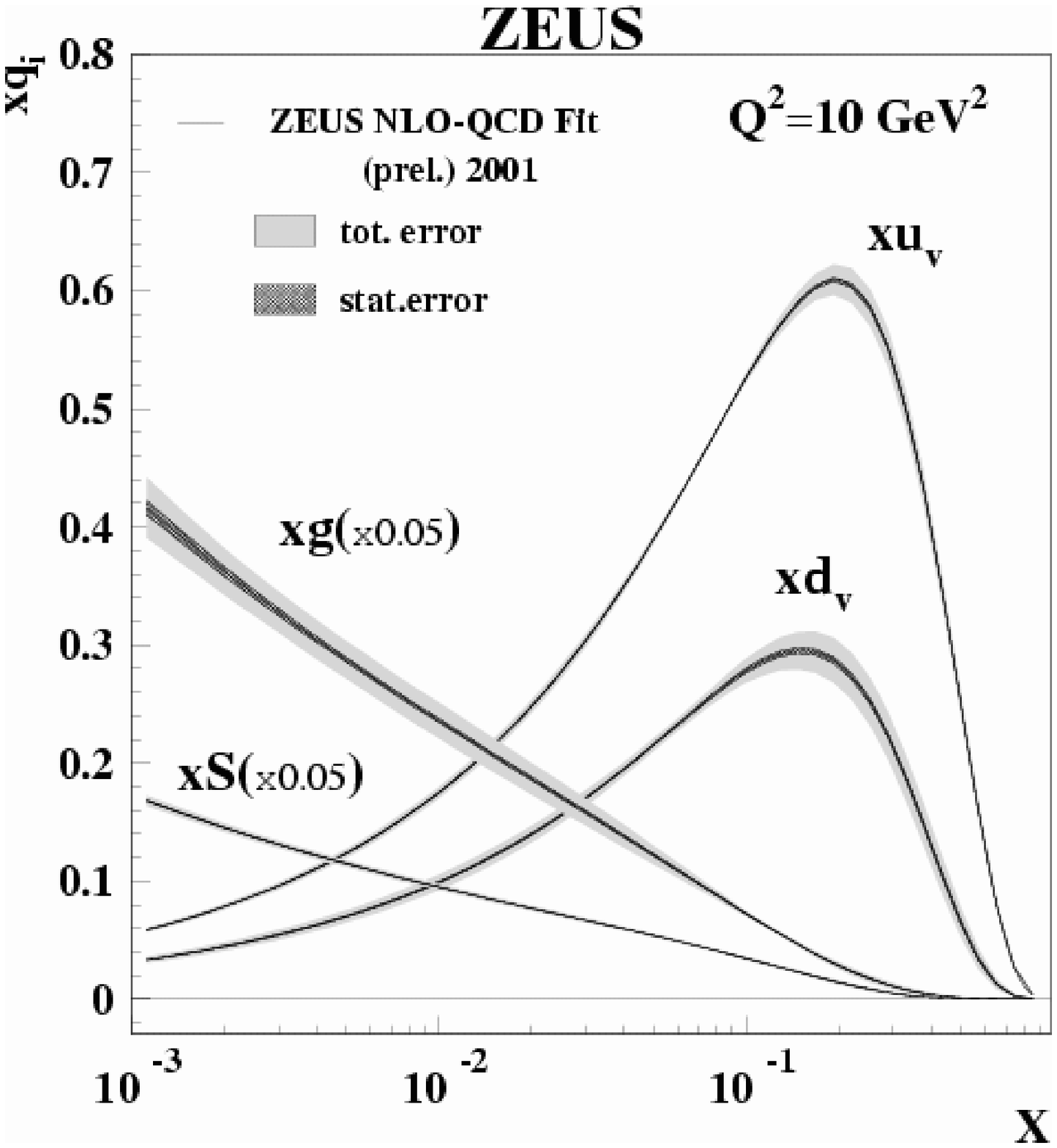,width=7.2cm}}
\end{picture}
\caption{a) Gluon distribution of the proton from H1,
b) parton distributions of the proton from ZEUS }
\label{fig:gluon}
\end{figure}

In Fig.~\ref{fig:gluon}a, the gluon distribution of the proton is shown
for $Q^2=5\,$GeV$^2$ from the H1 experiment \cite{f2h1}.
In the $x$ regions where the $Q^2$ lever arm is longest, $10^{-3}<x<10^{-2}$,
the precision is of order $10\%$.
The gluon distribution as determined from the ZEUS experiment 
\cite{alphaszeus}
is shown in Fig.~\ref{fig:gluon}b together with the other parton distributions
extracted from the data at $Q^2=10\,$GeV$^2$.

\subsubsection{\boldmath
Determination of the strong coupling constant $\alpha_s$}

\begin{figure}[htb]
\setlength{\unitlength}{1cm}
\begin{picture}(6.0,9)
\put(4,0){\epsfig{file=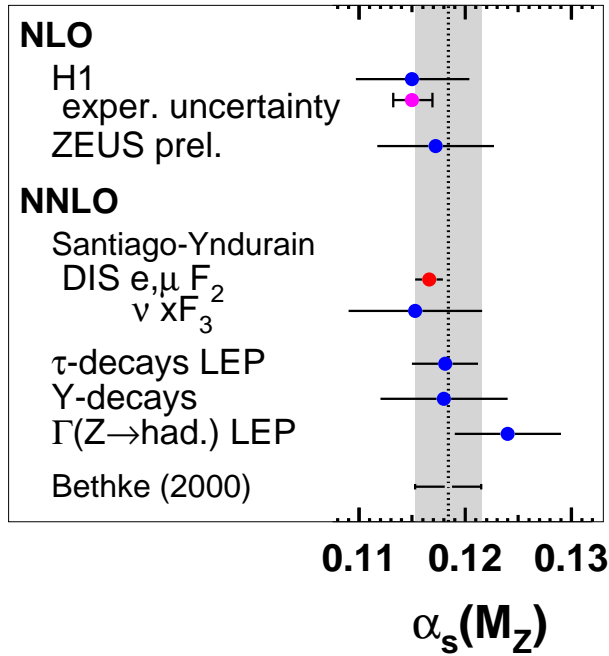,width=8cm}}
\end{picture}
\caption{\label{fig:alphas} Determinations of the strong coupling constant}
\end{figure}

\noindent
In Fig.~\ref{fig:alphas},
determinations of the strong coupling constant $\alpha_s$ 
are shown, grouped in NLO and NNLO QCD extractions. 
As a reference, the grey band shows the value given in a recent review of 
Bethke \cite{bethke}.
Both HERA experiments \cite{f2h1,alphaszeus} 
extract $\alpha_s$ at the NLO QCD level. 
Their $\alpha_s$ values are fully compatible with each other. 
The H1 experiment used their own data and fixed target data from proton 
target only, avoiding thereby target mass corrections and 
regions affected by higher twist effects. 
Their total experimental error is $0.002$, shown by the second 
symbol. 
The total error bar is dominated by the renormalisation 
scale uncertainty.
In this analysis, the scale was varied by a factor $2$ down and up. 

These theoretical uncertainties are expected to be much smaller when 
using higher order calculations. 
The next-to-next-to leading order (NNLO) corrections
are partly available \cite{moch}. 
The two-loop coefficient functions of $F_2,F_3$ and $F_L$ have been
calculated~\cite{vanNeerven:1991nn}, and have been
completely checked~\cite{Moch:1999eb}. 
For the three-loop anomalous dimensions $\gamma_{\rm pp}^{(2)}$, 
only partial results are available thus far. These include a finite
number of fixed Mellin
moments~\cite{Larin:1997wd,Retey:2000nq}, 
both for $F_2$ and $F_3$, the large $n_f$-limit~\cite{Gracey:1994nn} 
of $\gamma_{\rm qq}^{(2)}$ and $\gamma_{\rm gg}^{(2)}$, in the latter case 
only the coefficient of the colour factor $n_f^2 C_A$, and several terms 
relevant to the small-$x$ limit~\cite{Catani:1994sq}.

The corrections available so far are already sufficient to determine 
$\alpha_s$ at the NNLO QCD level by averaging the data in $Q^2$ bins
to calculate Mellin moments.
This approach has been followed by Santiago and Yndurain 
where the results \cite{yndurain} are shown in Fig.~\ref{fig:alphas}.
The error from the extraction using $ep$ and $\mu p$ data
is extremely competitive when compared to other 
determinations of $\alpha_s$. 
The central value and the experimental uncertainty should be confirmed
by other groups.
Regarding the theoretical uncertainties in $\alpha_s$ from 
deep inelastic scattering, van Neerven and Vogt also expect 
a theoretical error below $1\%$ \cite{vogt}. 
Determinations of $\alpha_s$ \cite{yndurain,kataev} using $\nu N$ data 
agree within errors which are, however, 
larger compared to the $ep,\mu p$ case
(Fig.~\ref{fig:alphas}).

It can be anticipated that 
the theoretical progress in the NNLO QCD corrections together 
with the increasing precision in the $F_2$ measurements will lead to
the most precise determinations of $\alpha_s$.

\subsection{Deep inelastic neutrino--nucleon scattering}

\noindent
The neutrino data cover a complementary kinematic region up to
$Q^2=250\,$GeV$^2$ and $0.01<x<1$.
These data are sensitive to the different quark flavours and are 
dominated by charged current interactions.

Recently the CCFR collaboration has re-analysed their data \cite{ccfr}
in the light of major improvements in the
theoretical understanding of the heavy quark contributions.
They changed the analysis method of extracting information on the
structure functions.
Their $F_2$ data are now well consistent with the charged lepton data
at the level of $5\%$, after correcting for the charge factor $18/5$,
and when using NLO QCD calculations of the heavy quark contributions.

\begin{figure}[htb]
\setlength{\unitlength}{1cm}
\begin{picture}(6.0,12)
\put(4,0){\epsfig{file=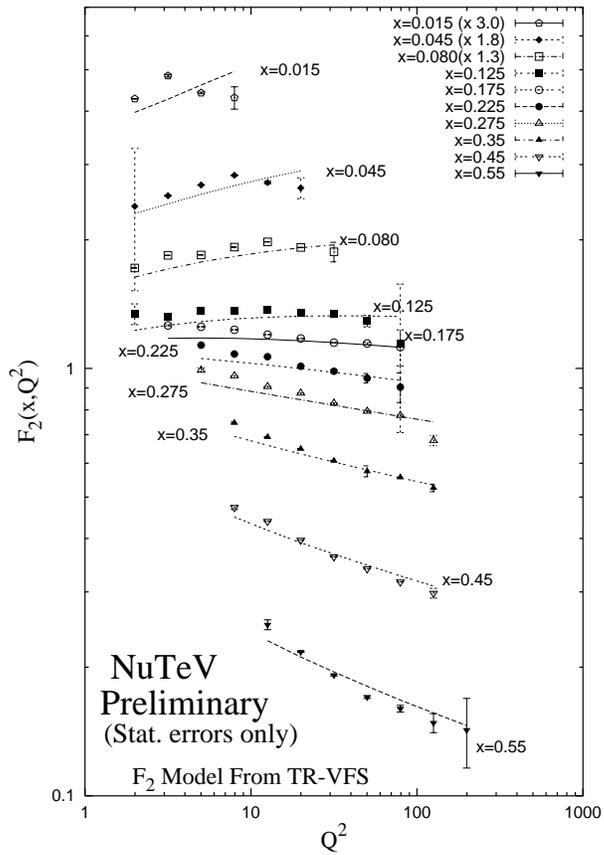,width=8cm}}
\end{picture}
\caption{Proton structure function $F_2$ from
neutrino--nucleon scattering}
\label{fig:neutrino} 
\end{figure}

The successor experiment \cite{naples}
NUTEV has overall much better experimental 
conditions.
The collaboration just released a first glance at their $F_2$ data so
far with statistical errors only (Fig.~\ref{fig:neutrino}).
While in this measurement, the data sets of the neutrino and anti-neutrino beams
have been combined, the clean separation of the $\nu$ and $\bar{\nu}$ beams
give them the potential for extracting momentum distributions 
for different quark flavours in the proton.

\subsection{Drell-Yan production of muon pairs}

\noindent
Measurements of the Drell-Yan production of muon pairs give access to 
the anti-quark distributions of the proton at high values of $x$  \cite{nusea}.
The ratio of proton--deuterium and proton--hydrogen target data is
directly related to the $\bar{d}/\bar{u}$ ratio.
For highly asymmetric kinematics of the fractional momenta of the 
incoming quark and anti-quark, $x_1 \gg x_2$, both of which can be 
calculated from the kinematics of the muon pair, the ratio is given by
\begin{equation}
\frac{\sigma(p d)}{2\;\sigma(p p)} \sim \frac{1}{2}
\left[1+ \frac{\bar{d}(x_2)}{\bar{u}(x_2)}\;\right] \; .
\end{equation}
This relation is valid for the assumption that the parton distributions
of the nucleon obey charge symmetry, $u_p(x)=d_n(x)$ etc.

\begin{figure}[htb]
\setlength{\unitlength}{1cm}
\begin{picture}(6.0,8)
\put(1.5,4){\epsfig{file=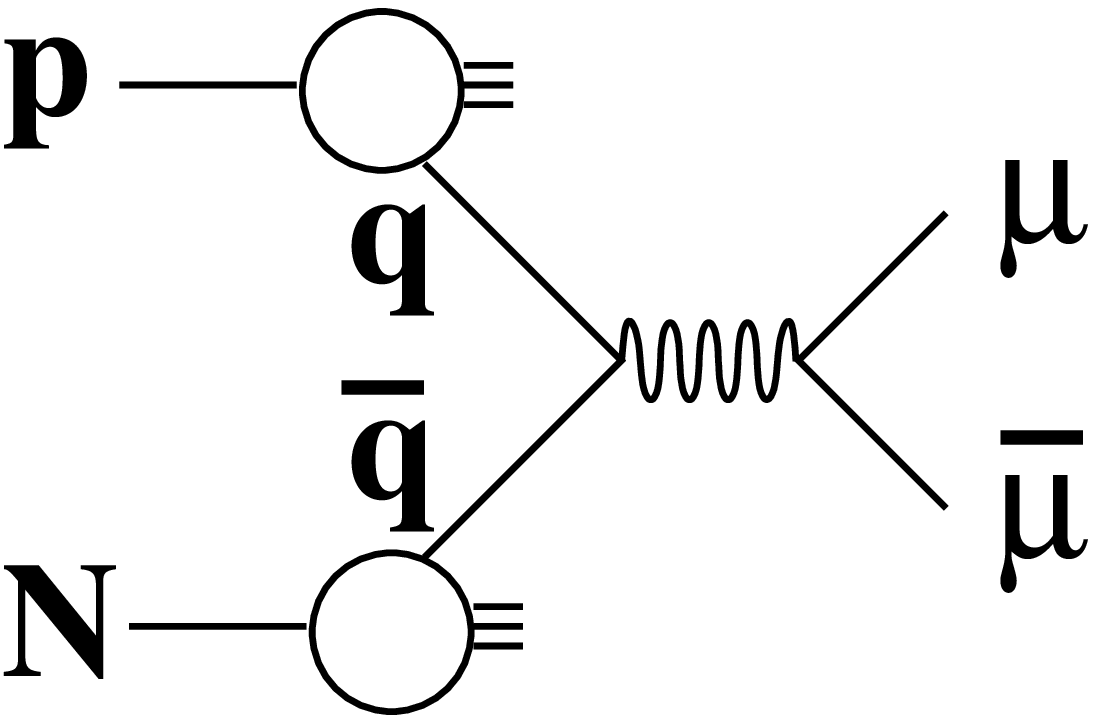,width=3cm}}
\put(6,0){\epsfig{file=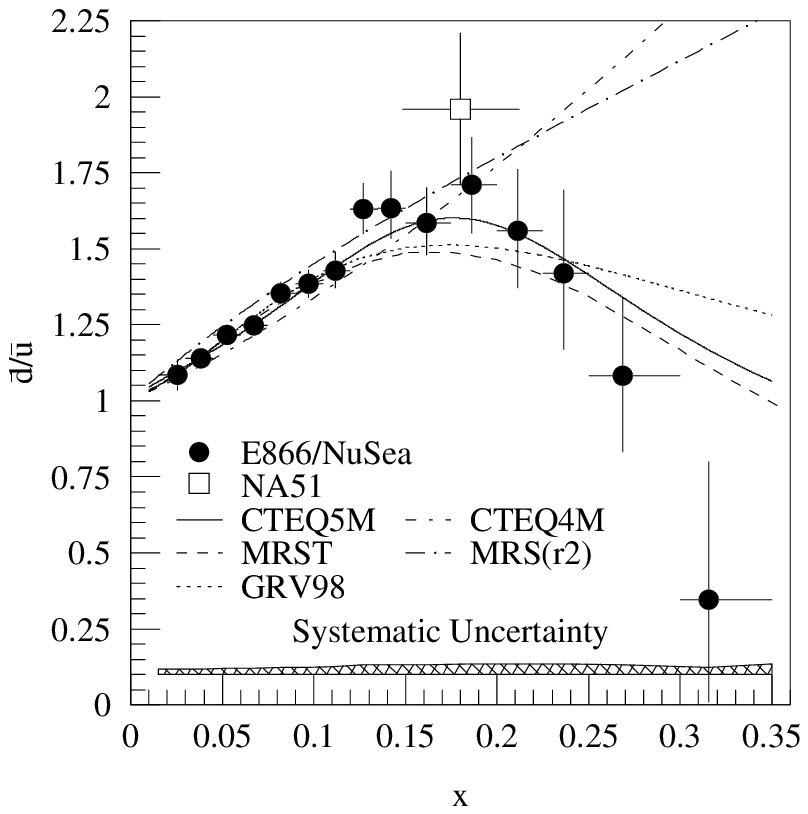,width=8cm}}
\end{picture}
\caption{ Ratio of the anti-down-- over
anti-up--quark distributions in the proton }
\label{fig:dbarubar}
\end{figure}

Fig.~\ref{fig:dbarubar} gives the measured ratio $\bar{d}/\bar{u}$ as a 
function of $x$ showing a large asymmetry which is likely to be of 
non-perturbative origin.

\subsection{Di-jet cross sections in hadron--hadron collisions}

\begin{figure}[htb]
\setlength{\unitlength}{1cm}
\begin{picture}(16.0,10)
\put(3,7.9){\epsfig{file=pp-pt2.eps,width=1.6cm}}
\put(0,0){\epsfig{file=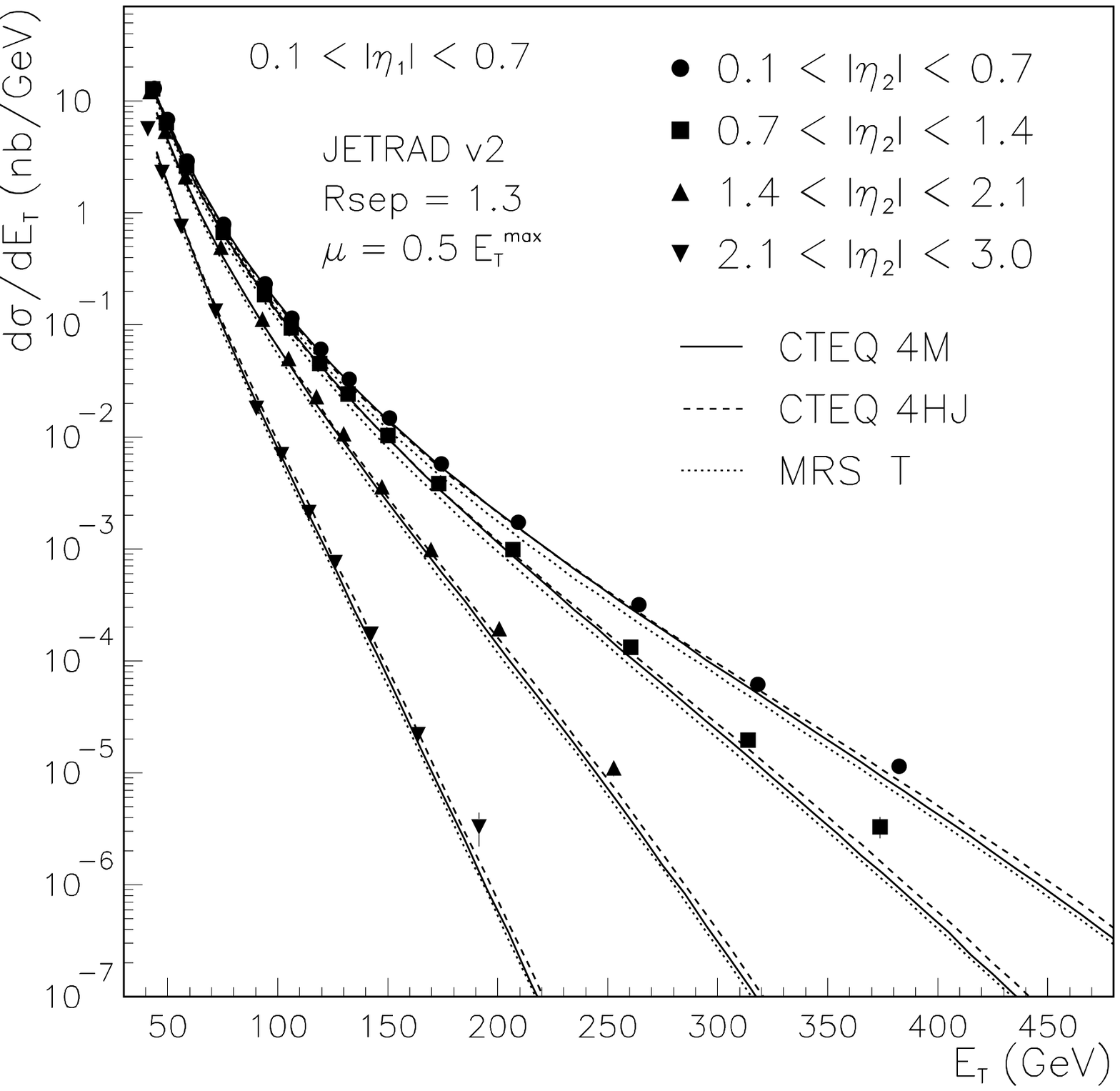,width=8.3cm}}
\put(6.,-3.6){\epsfig{file=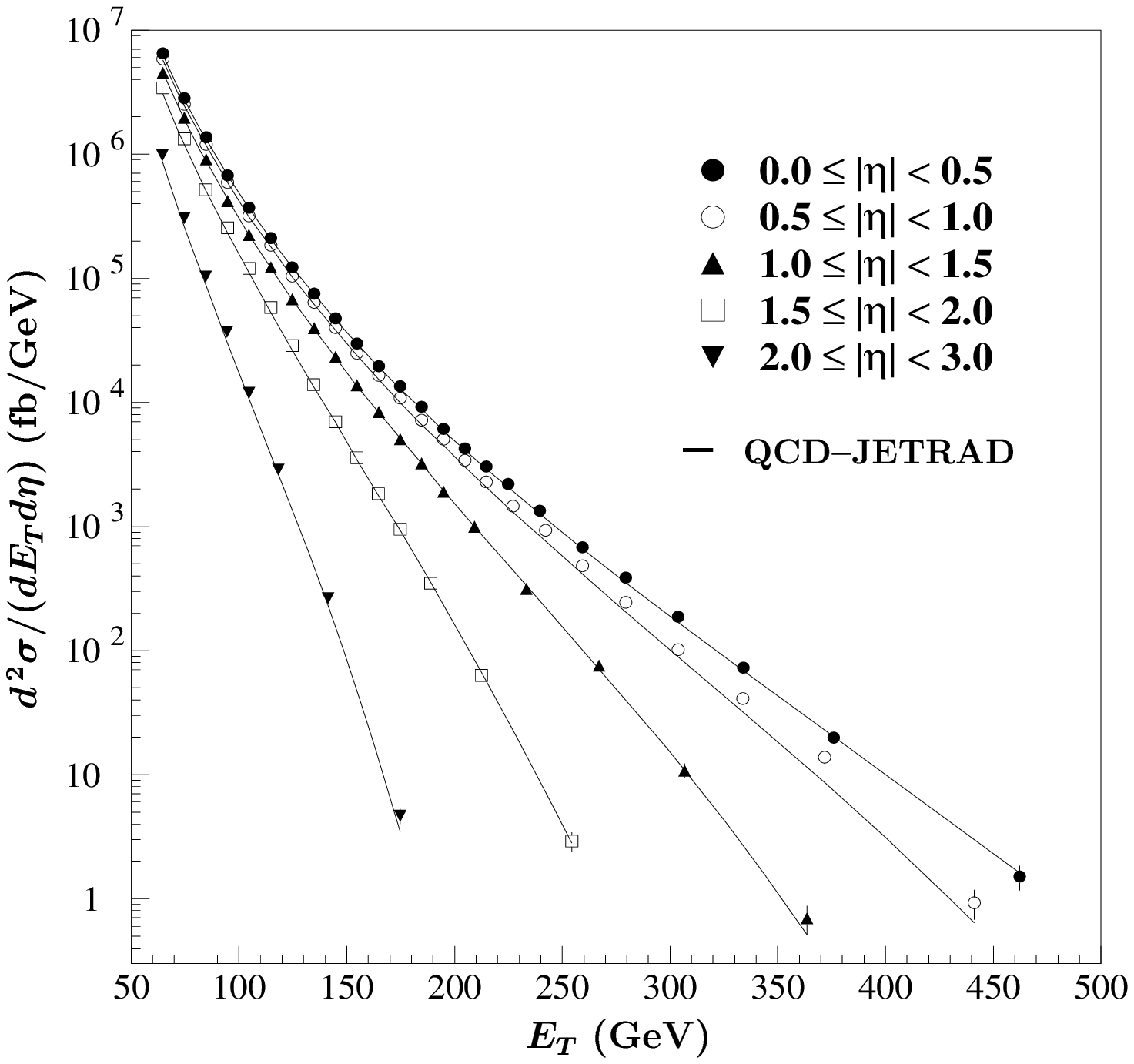,width=11.6cm}}
\end{picture}
\caption{ Differential di-jet cross section
in proton--anti-proton collisions from the CDF and D0 experiments}
\label{fig:cdf}
\end{figure}

\noindent
Jet measurements at the TEVATRON collider challenge the 
Standard Model at the smallest distance scales and simultaneously 
constrain the parton distributions of the proton.
Both collaborations CDF~\cite{cdf} and D0~\cite{dzero}
have published measurements of the 
di-jet cross sections in terms of the jet transverse energies 
$E_{T,i}$ and their pseudo-rapidities $\eta_i$ ($i=1,2$)
(Fig.~\ref{fig:cdf}).

In leading order QCD, these observables are related to the 
parton fractional momenta $x_{\pm}$ of the two incoming partons via
\begin{equation}
x_{\pm} = \frac{E_T}{\sqrt{s_{p\bar{p}}}} 
\left( \exp{(\pm \eta_1)}+\exp{(\pm \eta_2)}\right) \; .
\end{equation}
The data give access to the colour charge weighted sum of the quark and gluon 
distributions in the proton.
They cover the high-$Q^2$ -- high-$x$ region of Fig.~\ref{fig:kinematic}.
Here the quark distributions are rather well constrained so that the 
di-jet data give information on the high-$x$ gluon distribution of the proton.

The experiments express the sensitivity of their data
by comparing with NLO QCD
calculations using different sets of parton distributions of the proton 
obtained from global fits.
While the D0 collaboration reports a consistent description of their measurements
with some of the sets, the CDF collaboration finds that no set describes their data.
It is planned to include the di-jet data in global fits of the parton
distribution functions.
This enables studies of the high-$x$ gluon distribution of the proton,
and of the differences in the data of the two experiments.

\subsubsection{How good are predictions for cross sections involving
proton structure}

\noindent
A frequently asked question is, how well do we know the parton distributions
of the proton?
Recently, major progress has been achieved concerning the
correct treatment of the correlated experimental errors in global fits
\cite{f2h1,alphaszeus,pascaud,Botje:1999dj,Giele:1998gw,pumplin}.
These developments provide the tools to combine different data sets 
with a clear interpretation of the resulting $\chi^2$ per degree of freedom,
and allow error bands to be drawn for the parton distributions.
Examples are shown in Fig.~\ref{fig:gluon}.

To understand the impact of our knowledge of the proton structure
on cross section predictions, e.g. for the LHC, benchmark processes
are being studied and are directly related to the proton structure function
measurements.
For example, the $W$-boson production cross section is varied with respect to 
the incoming partons.
It is then checked whether these variations are within the uncertainties
of the structure function data.
Recent studies \cite{thorne,tung} 
indicate an accuracy of the $W$ production cross section at
the LHC of $2-4\%$.

\section{The Photon}

\noindent
Quantum fluctuations of the photon provide an ideal laboratory for
analysing the genesis process of hadronic structures.
The splitting of the photon into a quark--anti-quark pair 
at quark fractional momenta $0.1<x<1$ was already
rather well understood from comparisons of the measurements at
the PETRA and PEP $e^+e^-$ colliders with the theoretical predictions.
In contrast, the gluon developing in the hadronic 
fluctuations and the quark structure at low $x$ were essentially unknown before
HERA and LEP.

\subsection{Deep inelastic electron--photon scattering}

\begin{figure}[htb]
\setlength{\unitlength}{1cm}
\begin{picture}(6.0,6.5)
\put(0.5,2){\epsfig{file=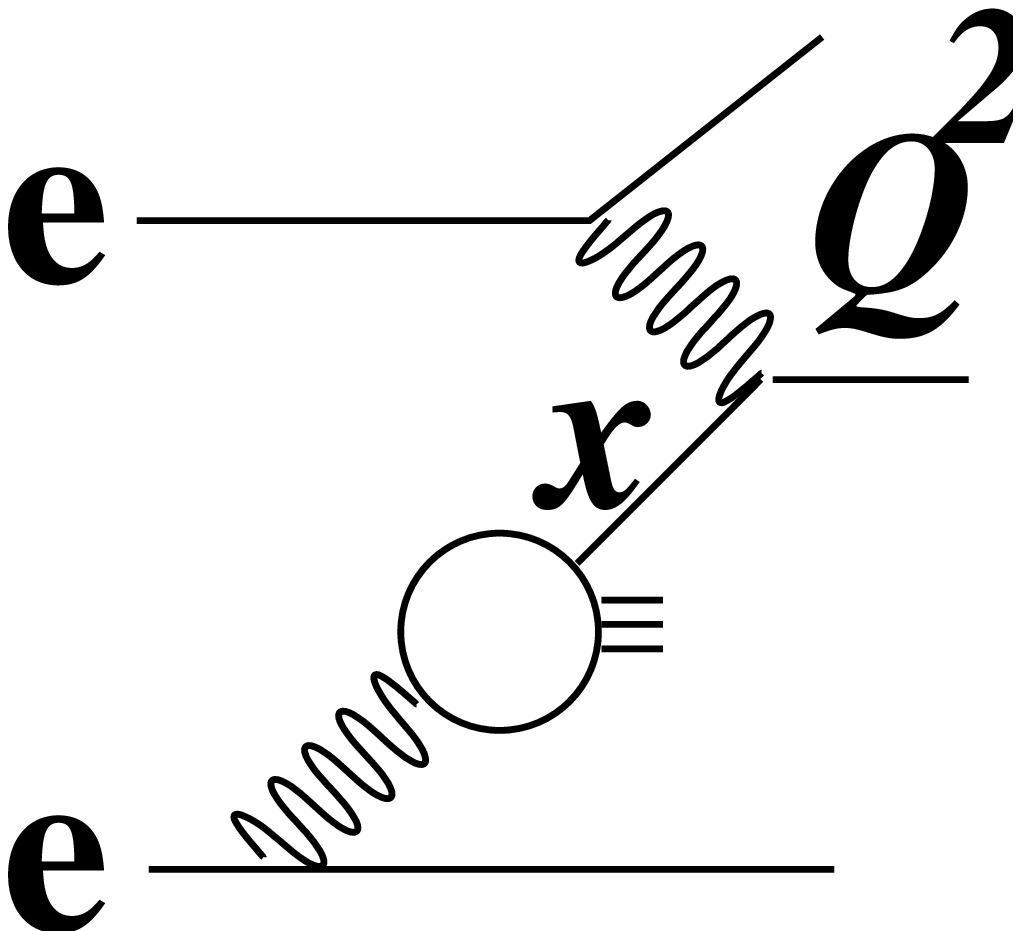,width=2.5cm}}
\put(4,0){\epsfig{file=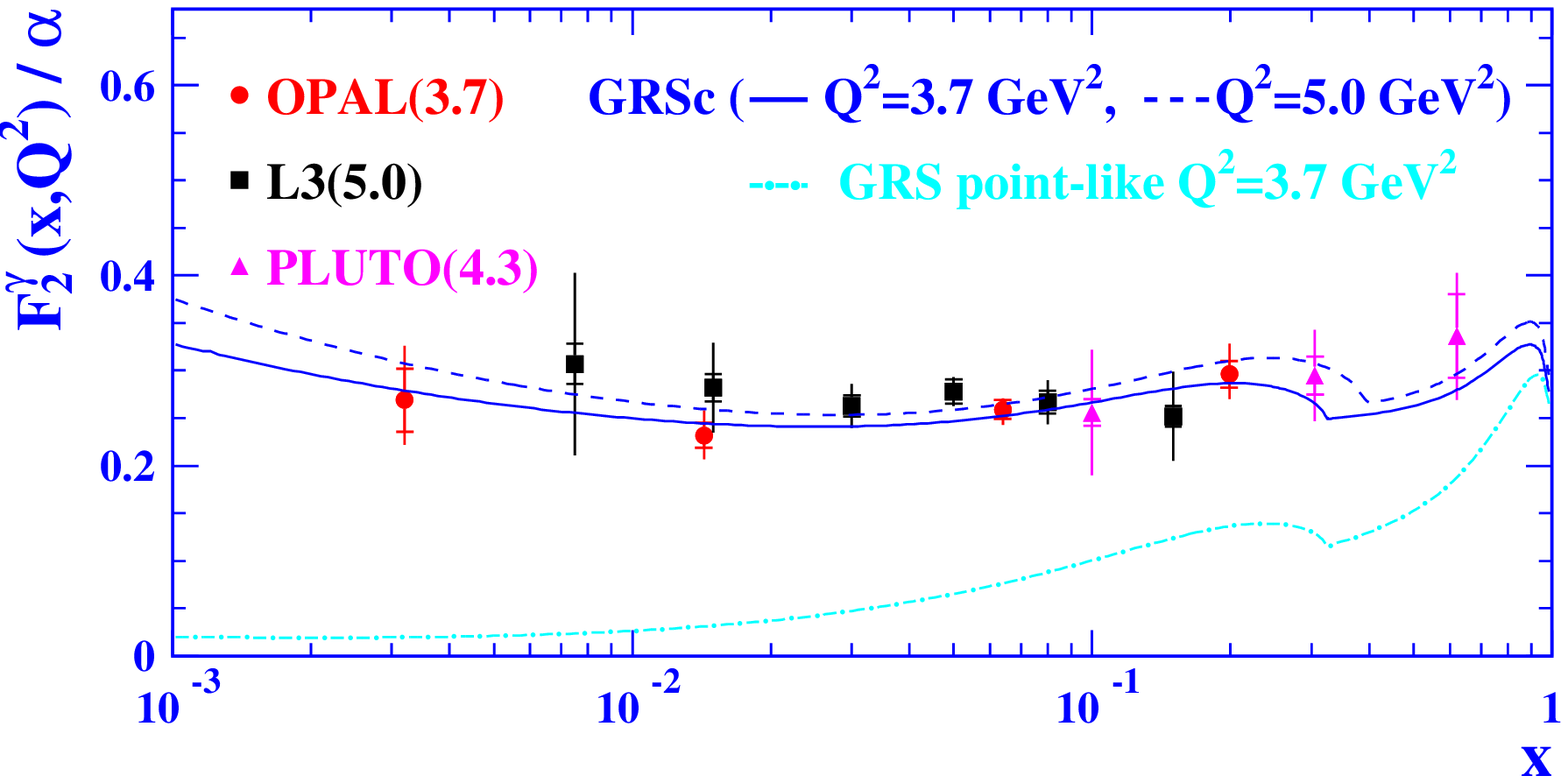,width=12cm}}
\end{picture}
\caption{\label{fig:photon-f2x} Photon structure function $F_2^\gamma/\alpha$ 
at low-$x$}
\end{figure}

\noindent
In Fig.~\ref{fig:photon-f2x}, a diagram of deep inelastic electron--photon
scattering is shown.
As for the proton case, the double differential cross section is related
to the photon structure function $F_2^\gamma$ by eqs.
~(\ref{eq:nc},\ref{eq:f2approx}).
The contributions of the structure function $F_L$ describing the 
structure of the transverse photon when probed by a longitudinal photon, 
and the contributions
of the $Z$ boson exchange are small in the kinematic region presently 
covered and are usually neglected in the analysis.

In Fig.~\ref{fig:photon-f2x}, measurements \cite{opal,l3,pluto} 
of $F_2^\gamma/\alpha$
are shown as a function of $x$ for {$3.7 < \langle Q^2\rangle < 5$ GeV$^2$}
(data comparison as in \cite{nisius}).
In the past few years, 
major improvements in the analysis techniques, especially at low-$x$,
have been achieved, giving an accuracy of $F_2^\gamma$ of the order of $10\%$.
The curves represent the predictions of Gl\"uck, Reya, and Schienbein
for the purely perturbative contribution, and separately the summed
perturbative and non-perturbative parts \cite{grs}.
The deviation of the data from the perturbative part indicates
the hadron-like contribution to the structure function data.

\begin{figure}[htb]
\setlength{\unitlength}{1cm}
\begin{picture}(14.0,12.)
\put(2,0){\epsfig{file=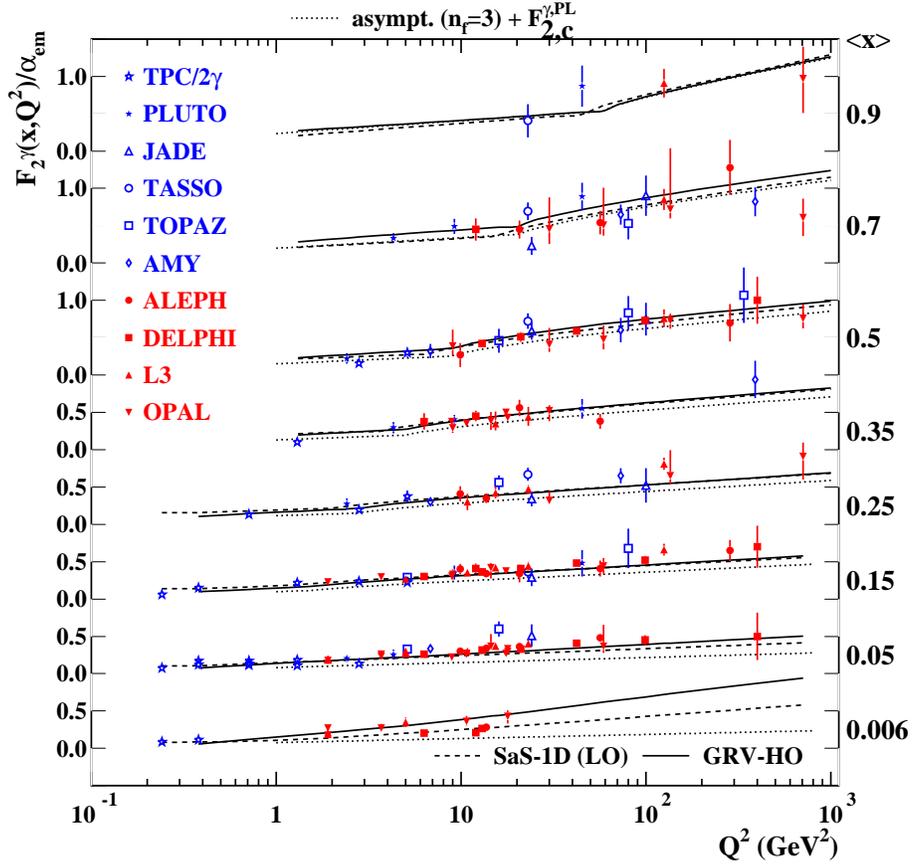,width=12cm}}
\end{picture}
\caption{\label{fig:photon-f2q2} Photon structure function $F_2^\gamma/\alpha$ }
\end{figure}

In Fig.~\ref{fig:photon-f2q2}, measurements of $F_2^\gamma/\alpha$
are summarized as a function of $Q^2$ in bins of $x$ 
(updated version from \cite{soldner}).
In comparison to the previous measurements of the photon structure,
the LEP data access the high-$Q^2$ region up to $10^3$ GeV$^2$,
and the low-$x$ regime down to $x\sim 10^{-3}$ for $Q^2\ge 2$ GeV$^2$.

\begin{figure}[htb]
\setlength{\unitlength}{1cm}
\begin{picture}(16.0,9)
\put(-0.5,0){\epsfig{file=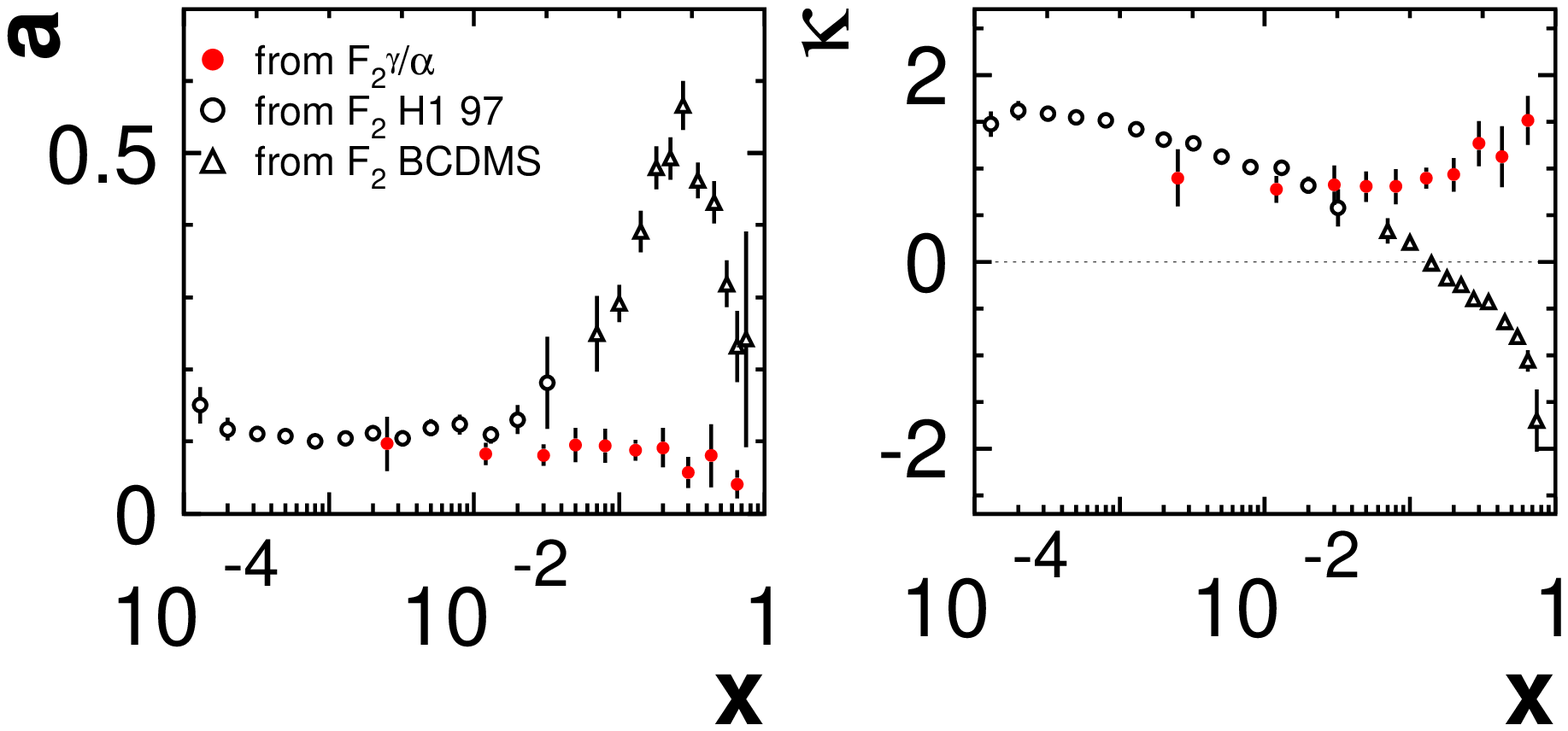,width=17cm}}
\put(14.0,3.){\epsfig{file=quark-gluon-split.eps,width=1.cm}}
\put(11.0,3.){\epsfig{file=gluon-split.eps,width=0.8cm}}
\put(14.3,8){\epsfig{file=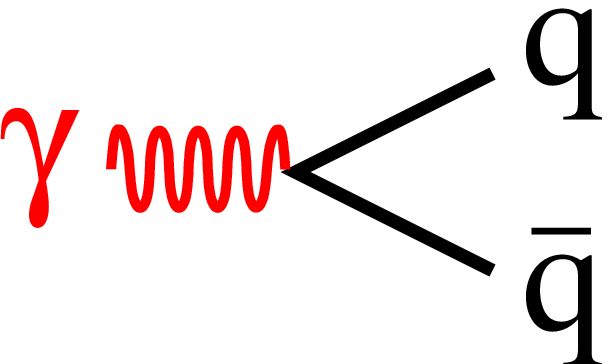,width=1.5cm}}
\end{picture}
\caption{\label{fig:photon-ak} Quark distribution of the photon 
at $Q^2=0.3$ GeV$^2$
and corresponding scaling violations (closed symbols) compared to
the proton (open symbols)}
\end{figure}

It is instructive to compare the photon data to the hadronic structure
of the proton.
In Fig.~\ref{fig:photon-ak}, the measurements of $F_2^\gamma/\alpha$
are analysed using the same description eq.~(\ref{eq:f2me}) 
as used above for the proton.
The photon shows no valence quark structure, as expected for this quantum fluctuation.

It is interesting to note that the values of the parameter $a$, 
that describes the quark distribution for $Q^2=0.3$ GeV$^2$,
are at the same level as those of the sea quarks in the proton.
At large values of $x>0.1$, the scaling violations $\kappa$ are expected 
to be of order $1$ from the splitting 
of the photon into a quark--anti-quark pair \cite{witten}.
The data are fully compatible with this prediction.
They strongly differ from the scaling violations observed in the measurements
of the proton structure.

A particular virtue of the photon measurements is their potential to
answer questions on universal properties of hadronic structures, 
e.g., do parton radiation processes lead to universal
hadronic structures at small values of $x$?
At low values of $x$, the proton data exhibit positive and increasing  
scaling violations from the contribution of the gluons.
The photon data now access this interesting domain of phase space;
however, the precision of the data is not yet good enough to decide
whether or not the scaling violations start to deviate from $1$ and
follow the scaling violations of the proton.

\subsection{Di-jet measurements from photon--proton and photon--photon collisions}

\begin{figure}[htb]
\setlength{\unitlength}{1cm}
\begin{picture}(6.0,11.)
\put(0.5,10){\large a)}
\put(9,10){\large b)}
\put(3.5,8){\epsfig{file=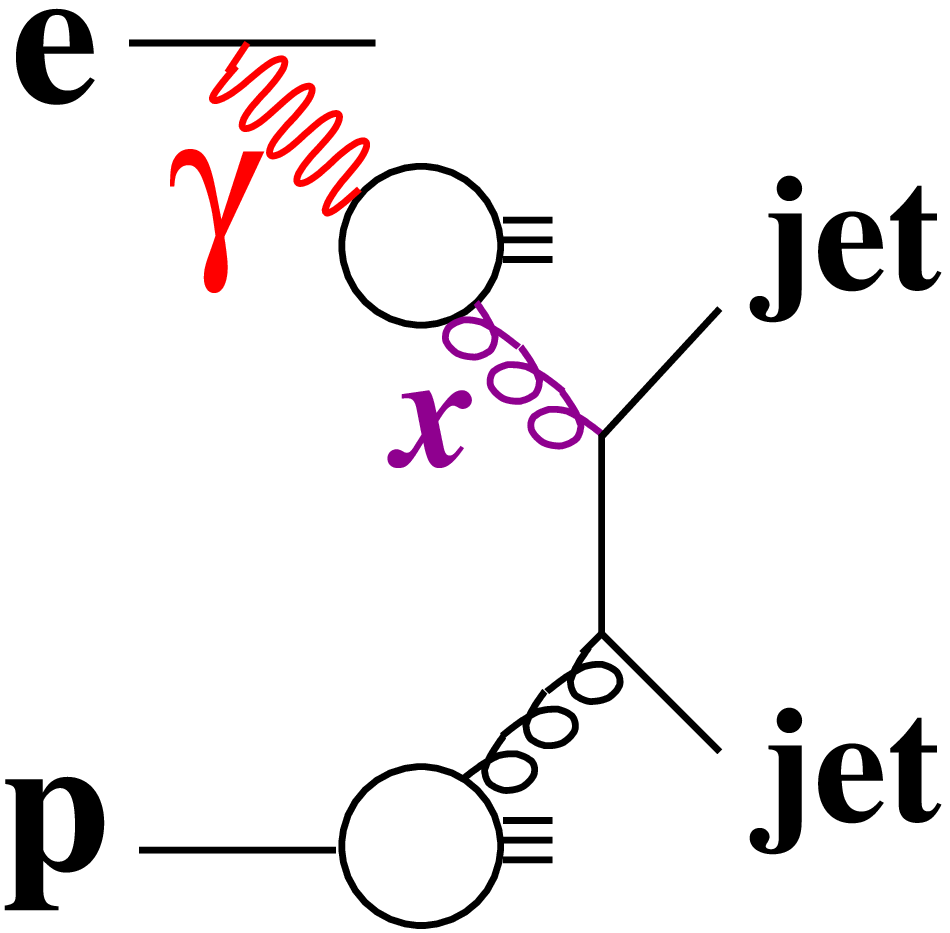,width=2.5cm}}
\put(-0.2,-0.2){\epsfig{file=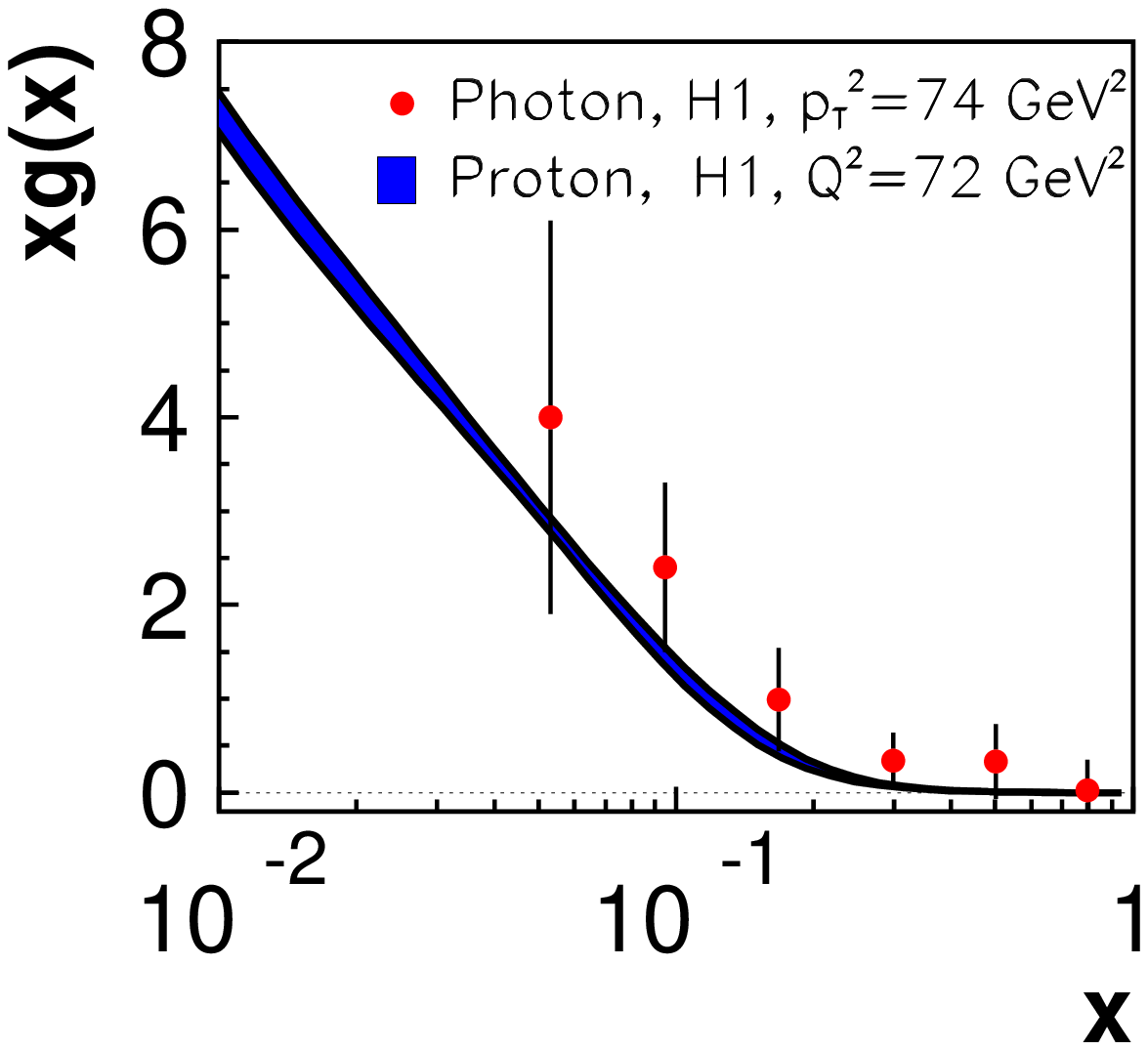,width=9cm}}
\put(11.5,8){\epsfig{file=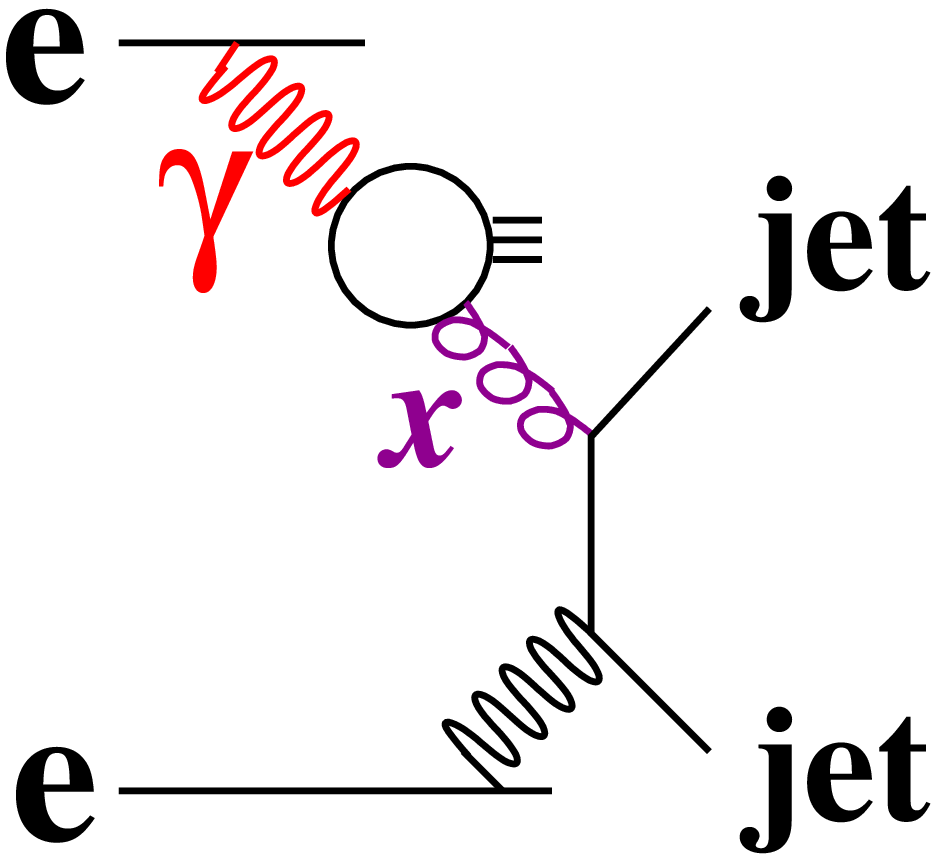,width=2.5cm}}
\put(8.9,0.5){\epsfig{file=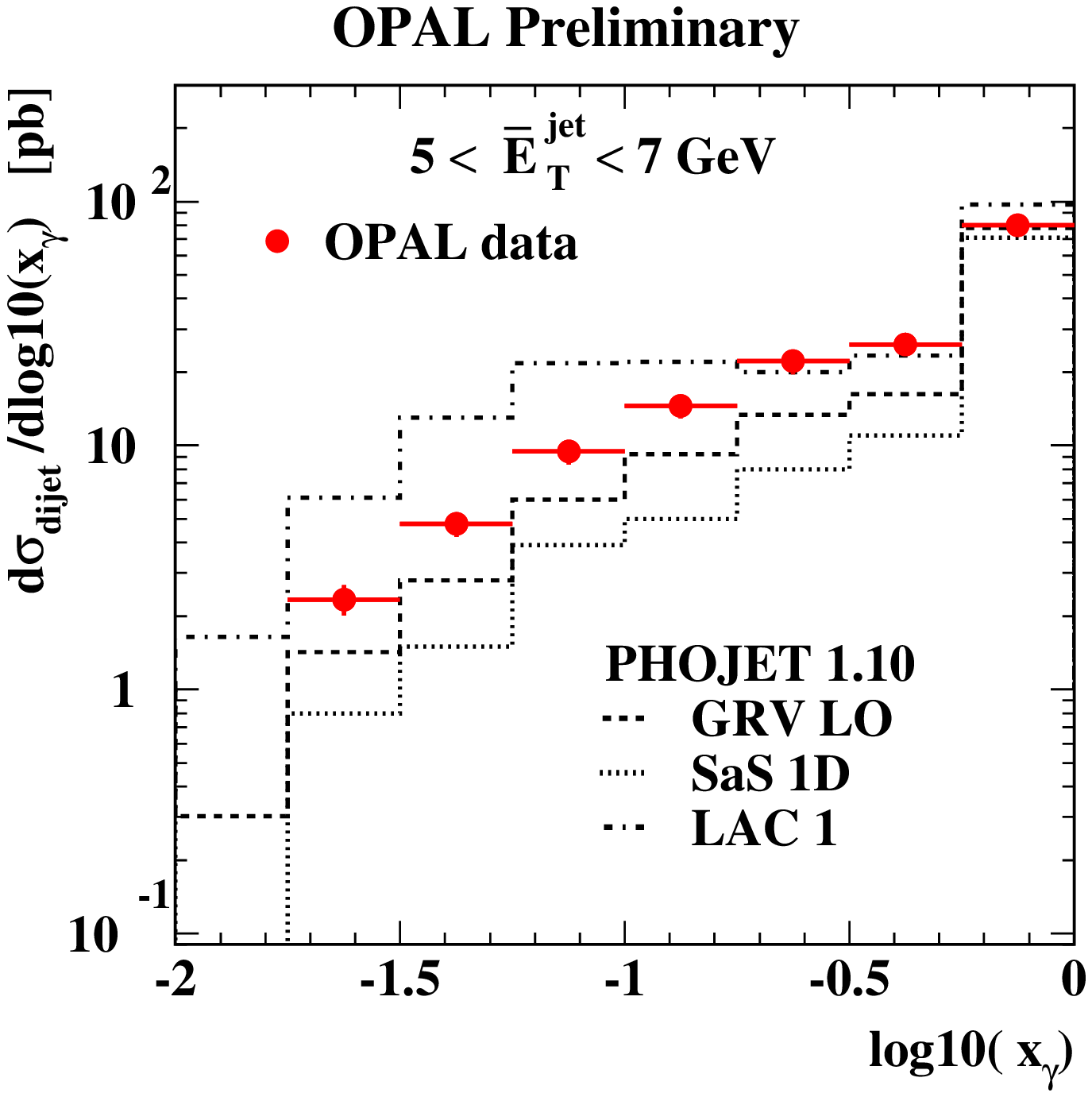,width=7.5cm}}
\end{picture}
\caption{\label{fig:photon-gluon} a) Gluon distribution of the photon from 
di-jet measurements in $\gamma p$ interactions in comparison to 
the gluon distribution of the proton,
b) di-jet cross section in $\gamma\gamma$ interactions compared to QCD calculations
using different input parton distributions for the photon}
\end{figure}

\noindent
While the deviation of the data from the purely perturbative prediction
in Fig.\ref{fig:photon-f2x} gives indirect evidence for a gluon contribution
to the hadronic structure, the first direct measurement of the gluon
distribution of the photon has been obtained in di-jet measurements at
HERA \cite{gammagluon}.
In Fig.\ref{fig:photon-gluon}a, a recent measurement from H1 
\cite{h1gluon} is shown in comparison to the gluon as measured in the
proton \cite{f2h1}.
The data are compared at the resolution scales $Q^2=p_t^2\sim 70$ GeV$^2$.
Although the photoproduction measurement is limited in precision owing
to underlying event effects, 
the similarity of the distributions possibly hints to a universal
gluon distribution in hadronic structures which are based on quarks.

Major improvements in the precision of the 
gluon distribution of the photon are expected to come from LEP data
where effects of the underlying event can be avoided, in principle.
In Fig.\ref{fig:photon-gluon}b, a preliminary OPAL measurement  \cite{wengler}
of the 
differential di-jet cross section is shown as a function of the 
logarithm of $x$.
The different histograms represent calculations of the cross section
using different input parton distributions, and demonstrate the high 
sensitivity of the data to the gluon contribution.

\section{Concluding remarks}

\noindent
Structure function measurements again become a field of precision 
data, this time covering a much larger phase space in $x$ and $Q^2$.
Together with new theoretical developments on 
NNLO QCD corrections (and beyond) to deep inelastic lepton--proton 
scattering, 
the data already imply a new level of precision for the value of 
the strong coupling constant.

Comparisons of the proton data with measurements on the photon 
structure allow universal properties of hadronic structures to
be studied directly from the measurements themselves.
The recent photon data cover the relevant phase space and
are in precision close to what is needed for firm statements.
 
\section*{Acknowledgments}
For discussions and kind help,
I wish to thank many members of the CERN, DESY, and FNAL experiments.
I am especially grateful to
I.~Bertram,
S.~Bethke,
A.~Cooper,
A.~Dubak,
E.~Elsen,
F.~Happacher, 
M.~Klein, 
M.~Moritz,
D.~Naples,
C.~Niebuhr, 
R.~Nisius, 
E.~Rizvi, 
P.~Schleper,
S.~S\"oldner-Rembold,
M.~Stanitzki,
W.~Tung,
P.~Valente,
R.~Wallny, 
T.~Wengler, 
J.~Whitmore,
H.~Wieber, 
R.~Yoshida,
and the Rome Team!
I wish to thank S.~Moch for his kind advice concerning the NNLO developments
in deep inelastic scattering.
I wish to further thank Th.~M\"uller and the IEKP group of the University
Karlsruhe for their hospitality, and the Deutsche Forschungsgemeinschaft 
for the Heisenberg Fellowship.

\end{document}